\documentclass[a4paper,11pt]{article}
\pdfoutput=1 
\PassOptionsToPackage{dvipsnames}{xcolor}
\usepackage{jcappub} 

\usepackage[T1]{fontenc} 
\usepackage{tikz}
\usepackage[dvipsnames]{xcolor}
\usepackage{subcaption}
\usepackage{caption}
\usepackage{adjustbox}
\usepackage{svg}
\usepackage{fontawesome}
\usepackage{wrapfig}
\usepackage[normalem]{ulem}
\usepackage{cancel}

\usetikzlibrary{decorations.text}
\usetikzlibrary{calc}
\usetikzlibrary{angles,quotes}
\definecolor{saetass_yellow}{RGB}{243,172,75}
\newcommand{\FTdir}{}
\def\FTdir(#1,#2,#3){%
  \FTfile(#1,{{\color{saetass_yellow}\faFolderOpen}\hspace{0.2em}#3})
  (tmp.west)++(0.8em,-0.4em)node(#2){}
  (tmp.west)++(1.5em,0)
  ++(0,-1.3em) 
}

\newcommand{\FTfile}{}
\def\FTfile(#1,#2){%
  node(tmp){}
  (#1|-tmp)++(0.6em,0)
  node(tmp)[anchor=west,black]{\tt #2}
  (#1)|-(tmp.west)
  ++(0,-1.2em) 
}

\newcommand{\FTroot}{}
\def\FTroot{tmp.west}

\newcommand{\chng}[2]{%
  \ifmmode
    {\color{red}\cancel{#1}}%
    {\color{green}#2}%
  \else
    {\color{red}\sout{#1}}%
    {\color{green}#2}%
  \fi
}

\title{\texttt{SAETASS}: Solver for Astroparticle Equation of Transport Analysis in Spherical Symmetry}

\author[a]{J.M. García-Morillo,}
\author[a]{S. Menchiari}
\author[a]{and R. López-Coto}

\affiliation[a]{Instituto de Astrofísica de Andalucía (IAA-CSIC).\\18008, Granada. Spain.}

\emailAdd{jmorillo@iaa.es}
\emailAdd{smenchiari@iaa.es}
\emailAdd{rlopezcoto@iaa.es}

\abstract{In order to model astrophysical environments characterized by radial stratification, such as stellar winds, supernova remnants or expanding superbubbles; correctly understanding the transport of non-thermal particles in astrophysical plasmas is essential. While large-scale Galactic propagation codes exist, they are often optimized for Cartesian or cylindrical geometries and lack the efficiency or accessibility required for one-dimensional spherically symmetric problems. In this work, we present \texttt{SAETASS} (Solver for Astroparticle Equation of Transport Analysis in Spherical Symmetry), a novel, open-source numerical tool designed to solve the time-dependent transport equation for astroparticles. 

The solver is built upon a conservative finite-volume framework that ensures exact particle conservation and numerical stability. To manage the interplay between diverse physical processes, \texttt{SAETASS} employs a modular operator-splitting architecture. Radial advection and continuous momentum losses are treated using a second-order, shock-capturing MUSCL-Hancock scheme, while the diffusive operator is integrated via an implicit, batched Crank-Nicolson algorithm. This approach allows for the robust handling of steep gradients, spatial discontinuities and regularity conditions at the origin.

We rigorously validate the code through a suite of tests, comparing numerical results against analytical solutions for pure advection, diffusion and losses. Furthermore, the method of manufactured solutions is used to validate the non-stationary solutions. Finally, we demonstrate the solver's capabilities by modelling cosmic-ray proton transport in a real astrophysical scenario. Our results successfully recover established steady-state limits while revealing relevant pre-equilibrium temporal dynamics across Kolmogorov, Kraichnan and Bohm diffusion regimes. \texttt{SAETASS} provides the community with a lightweight, flexible tool for investigating particle acceleration and propagation in complex, radially dependent astrophysical environments.}

\begin{document}
\maketitle
\flushbottom

\section{The numerical study of astroparticle transport}
\label{sec:intro}

Understanding the transport of non-thermal particles in astrophysical plasmas is a central problem across a wide range of environments, from Galactic scales down to the vicinity of astrophysical particle accelerators. Over recent decades, increasingly detailed observations of non-thermal emissions have highlighted the need for accurate modelling tools for cosmic ray transport. Because the observed emissions arise from a complex interplay between particle acceleration, transport and energy losses, such models are essential for disentangling these effects and, thus, constraining the physics of the underlying acceleration processes \citep{Strong2007, Grenier2015}.

Fundamentally, the evolution of non-thermal particles is governed by kinetic theory in a six-dimensional phase space. However, in most astrophysical plasmas, frequent pitch-angle scattering induced by magnetic turbulence operates on timescales significantly shorter than those of macroscopic dynamical processes. This rapid scattering effectively isotropizes the particle velocities. As a result, the full directional phase-space distribution, $f(t,\mathbf{r},\mathbf{p})$, can be well approximated by a dominant isotropic term and a small anisotropic perturbation, which drives the spatial diffusive flux. By averaging the underlying kinetic equation over all momentum directions, the angular dependencies are integrated out. Consequently, the evolution of this nearly isotropic distribution function, $f(t,\mathbf{r},p)$, which now depends solely on the scalar momentum magnitude, under the action of advection, diffusion and momentum losses is commonly described by the transport equation \citep{Parker1965},
\begin{equation}
\label{eq:general_transport_equation}
    \frac{\partial f}{\partial t} 
    + \nabla \cdot \left( \mathbf{u}_\mathrm{w} f \right)
    +\frac{\partial}{\partial p}\left( \dot{p} f \right)
    = \nabla \cdot \left( D \nabla f \right)
    + Q,
\end{equation}
where $\mathbf{u}_\mathrm{w}(t,\mathbf{r},p)$ denotes the advective velocity of the background plasma, $D(t,\mathbf{r},p)$ the spatial diffusion coefficient, $\dot{p}(t,\mathbf{r},p)$ the momentum change rate and $Q(t,\mathbf{r},p)$ a generic term describing typically a source component, but that could also account for sink terms. We note that it is common to also include reacceleration processes in this modellisation, which is done by means of a diffusive term in momentum. However, we are not interested in this description. Equation~\eqref{eq:general_transport_equation} couples multiple physical mechanisms operating simultaneously across a broad range of spatial and temporal scales; its solution therefore requires dedicated numerical techniques, especially in coordinate systems, such as the spherical, which introduce non-trivial specificities.

In many astrophysical systems of interest, the physical configurations exhibit a dominant radial dependence. This is the case, for instance, in stellar winds, supernova-driven flows, expanding hot bubbles or spherically stratified halos around compact objects. In such settings, spherical symmetry emerges not merely as a simplification, but as the most physically consistent geometry for capturing the radial evolution of relevant quantities. Consequently, these systems are often modelled as stratified structures composed of concentric shells, where physical parameters such as magnetic field strength, gas density and temperature vary primarily with the radial coordinate. 

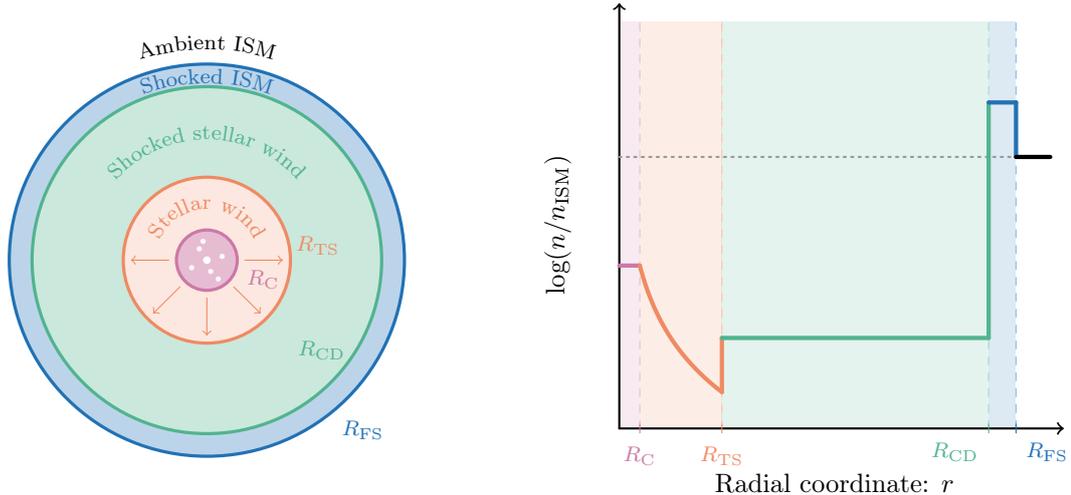
\begin{figure}
    \centering
    \begin{subfigure}[t]{0.49\linewidth}
        \centering
        \begin{tikzpicture}[line join=round, line cap=round]

\definecolor{C1}{RGB}{33,113,181}   
\definecolor{C2}{RGB}{239,138,98}   
\definecolor{C3}{RGB}{80,180,142}  
\definecolor{C4}{RGB}{204,121,167}  


\fill[white] (-3.2,-3.2) rectangle (3.2,3.2);

\filldraw[fill=C1!30, draw=C1, line width=1.2pt] (0,0) circle (2.6);

\filldraw[fill=C3!30, draw=C3, line width=1.2pt] (0,0) circle (2.3);

\filldraw[fill=C2!20, draw=C2, line width=1.2pt] (0,0) circle (1.1);

\filldraw[fill=C4!50, draw=C4, line width=1.2pt] 
    (0,0) circle (0.4);

\fill[white] (0,0) circle (0.05);
\foreach \pos in {(-0.1,0.15), (0.05,-0.15), (0.2,0.05), (-0.2,-0.1), (-0.05,0.25), (0.15,-0.25)}
    \fill[white, opacity=0.9] \pos circle (0.035);


\path[decorate, decoration={text along path, text={|\scriptsize\color{black}|Ambient ISM}, text align=center}] 
    (150:2.8) arc (150:30:2.8);

\path[decorate, decoration={text along path, text={|\scriptsize\color{C1}|Shocked ISM}, text align=center}] 
    (150:2.35) arc (150:30:2.35);

\path[decorate, decoration={text along path, text={|\scriptsize\color{C3}|Shocked stellar wind}, text align=center}] 
    (150:1.6) arc (150:30:1.6);

\path[decorate, decoration={text along path, text={|\scriptsize\color{C2}|Stellar wind}, text align=center}] 
    (150:0.68) arc (150:30:0.68);
\foreach \angle in {180, 225, 270, 315, 360}
    \draw[->, thin, C2] (\angle:0.5) -- (\angle:1.0);


\begin{scope}[rotate=0]
    \node[anchor=south west, font=\scriptsize, inner sep=2pt, text=C4] at (0.45, -0.45) {$R_\text{C}$};
    
    \node[anchor=south west, font=\scriptsize, inner sep=2pt, text=C2] at (1.1,0) {$R_\text{TS}$};
    
    \node[anchor=north east, font=\scriptsize, inner sep=2pt, text=C3] at (1.9, -1) {$R_\text{CD}$};
    
    \node[anchor=south west, font=\scriptsize, inner sep=2pt, text=C1] at (1.7, -2.45) {$R_\text{FS}$};
\end{scope}

\end{tikzpicture}
        \caption{Schematic representation of the stratified structure of a wind-blown bubble following the model in \citep{Weaver1977} model. The diagram identifies the central cluster core ($R_\text{C}$), the termination shock ($R_\text{TS}$), the contact discontinuity ($R_\text{CD}$) and the forward shock ($R_\text{FS}$), delimiting the different hydrodynamical zones of the system.}
        \label{fig:weaver_bubble}
    \end{subfigure}
    \hfill
    \begin{subfigure}[t]{0.49\linewidth}
        \centering
        \begin{tikzpicture}[line join=round, line cap=round, x=0.9cm, y=1.2cm]

\definecolor{C1}{RGB}{33,113,181}   
\definecolor{C2}{RGB}{239,138,98}   
\definecolor{C3}{RGB}{80,180,142}  
\definecolor{C4}{RGB}{204,121,167}  

\def\RcVal{0.3}  
\def\RsVal{1.5} 
\def\RcdVal{5.4} 
\def\RfsVal{5.8} 
\def\Rmax{6.3}   


\begin{scope}
    \clip (0,-3) rectangle (\Rmax, 1.5);

    \fill[C4, opacity=0.15] (0, -3) rectangle (\RcVal, 1.5);
    \fill[C2, opacity=0.15] (\RcVal, -3) rectangle (\RsVal, 1.5);
    \fill[C3, opacity=0.15] (\RsVal, -3) rectangle (\RcdVal, 1.5);
    \fill[C1, opacity=0.15] (\RcdVal, -3) rectangle (\RfsVal, 1.5);
    \fill[white] (\RfsVal, -3) rectangle (\Rmax, 1.5); 

    \draw[dashed, C4!50] (\RcVal, -3) -- (\RcVal, 1.5);
    \draw[dashed, C2!50] (\RsVal, -3) -- (\RsVal, 1.5);
    \draw[dashed, C3!50] (\RcdVal, -3) -- (\RcdVal, 1.5);
    \draw[dashed, C1!50] (\RfsVal, -3) -- (\RfsVal, 1.5);
\end{scope}


\draw[thick, ->] (0,-3) -- (\Rmax+0.2, -3); 
\node[font=\small] at ({\Rmax/2}, -3.6) {Radial coordinate: $r$}; 

\draw[thick, ->] (0,-3) -- (0, 1.7); 
\node[rotate=90, font=\small] at (-0.9, -0.75) {$\log(n/n_{\text{ISM}})$};

\draw[dotted, black!40, thick] (0,0) -- (\Rmax, 0);


\draw[thin, C4] (\RcVal, -2.95) -- (\RcVal, -3.05) node[below=1.5pt, font=\scriptsize] {$R_\text{C}$};
\draw[thin, C2] (\RsVal, -2.95) -- (\RsVal, -3.05) node[below=1.5pt, font=\scriptsize] {$R_\text{TS}$};
\draw[thin, C3] (\RcdVal, -2.95) -- (\RcdVal, -3.05) node[below left, font=\scriptsize] {$R_\text{CD}$};
\draw[thin, C1] (\RfsVal, -2.95) -- (\RfsVal, -3.05) node[below right, font=\scriptsize] {$R_\text{FS}$};


\draw[ultra thick, C4] (0, -1.2) -- (\RcVal, -1.2);

\draw[ultra thick, C2] plot[domain=\RcVal:\RsVal, samples=100] (\x, {-1.2 - 2*log10(\x/\RcVal)});

\draw[ultra thick, C2] (\RsVal, {-1.2 - 2*log10(\RsVal/\RcVal)}) -- (\RsVal, -2);

\draw[ultra thick, C3] (\RsVal, -2) -- (\RcdVal, -2);

\draw[ultra thick, C3] (\RcdVal, -2) -- (\RcdVal, 0.602);

\draw[ultra thick, C1] (\RcdVal, 0.602) -- (\RfsVal, 0.602);

\draw[ultra thick, C1] (\RfsVal, 0.602) -- (\RfsVal, 0);

\draw[ultra thick, black] (\RfsVal, 0) -- (\Rmax, 0);

\end{tikzpicture}
        \caption{Radial density profile $\log_{10}(n/n_\text{ISM})$ as a function of the radial coordinate. The profile illustrates the characteristic expansion in the unshocked wind region and the density jumps at the shock fronts, defining different zones of physical properties.}
        \label{fig:weaver_profile}
    \end{subfigure}

    \caption{Hydrodynamical structure of a prototypical stellar-wind bubble based on the framework given in \citep{Weaver1977}. This configuration serves as the benchmark for current transport analysis, where the radial gradients and localized shocks are more optimally treated through a dedicated numerical scheme in spherical coordinates.}
    \label{fig:weaver_intro_fig}
\end{figure}

A prototypical example of this approach is the self-similar solution for wind-blown bubbles proposed in \citep{Weaver1977}. This model describes the hydrodynamic evolution of massive stars winds and their interaction with the interstellar medium (ISM) around. The model identifies three distinct zones: a freely expanding stellar wind, a region of shocked wind and a swept-up shell of ambient gas. Additionally, in order to prevent singularities at the origin, one can define another region, the cluster core, not included in the original work. This framework, illustrated in Figure~\ref{fig:weaver_intro_fig}, remains the main tool for interpreting observations of hot bubbles around young stellar clusters, which have recently gained attention as potential PeVatrons \citep{Bykov2018, Aharonian2019, Morlino2021, Vieu2023}. However, while the hydrodynamical structure of these models is well established, the transport of cosmic rays within them has traditionally been treated with significant simplifications.

Indeed, existing models of particle distributions in environments like stellar-cluster wind bubbles often rely on analytical or semi-analytical approximations. These treatments frequently assume steady-state conditions or utilize one-zone approximations that, while potentially including energy losses, neglect the full time-radial dependence of the transport process \citep{Morlino2021}. No fully time-dependent calculation that integrates realistic loss processes with radially varying advection and diffusion has been widely established for these configurations. This is particularly relevant in expanding supernova-driven bubbles or wind-blown cavities created by stellar winds, where cooling, transport and injection mechanisms interplay in the dynamical evolution of the system.

This lack of analytical models is not particular to spherically symmetric systems and, thus, has led to the development of diverse numerical tools with the aim of solving the transport equation. Codes such as \texttt{GALPROP} \citep{Strong1998}, \texttt{DRAGON} \citep{Evoli2017}, \texttt{PICARD} \citep{Kissmann2014} and \texttt{USINE} \citep{Maurin2020} serve as indispensable tools for predicting large-scale distributions and diffuse radiation in the Milky Way. However, these codes are inherently tailored to Galactic-scale applications, relying on cylindrical or Cartesian geometries. Their internal structures are optimized for large-scale contexts and do not efficiently support spherically symmetric problems. Other options, as \texttt{RATPaC} \citep{Telezhinsky2012, Telezhinsky2013, Brose2016, Sushch2018, Brose2019}, are specifically designed for the study of supernova remnants in spherical symmetry but are not publicly available or open source. This lack of accessibility  prevents the community from adapting the code to a broader range of astrophysical environments, such as stellar winds or stratified circumstellar halos.

On the other hand, while more general magnetohydrodynamic codes like \texttt{Athena++} \citep{Stone2020} have incorporated cosmic-ray transport modules, these are typically part of complex, multi-dimensional architectures designed for high-performance computing on large clusters.

Furthermore, it is important to highlight a complementary approach to the Eulerian grid-based solvers: the resolution of the transport equation via stochastic differential equations (SDEs). This mathematical framework has been successfully implemented in recent developments of the \texttt{CRPropa} code \citep{AlvesBatista2022, Merten2025}, as well as in other dedicated studies \citep{Kopp2012, Krumholz2022}. These Lagrangian, particle-tracking methods are indeed powerful for capturing fully time-dependent problems, high-dimensional phase spaces or complex magnetic topologies. Nevertheless, Eulerian grid-based methods remain highly desirable, and often superior, when macroscopic symmetries can be exploited. Specifically, under assumptions such as spherical symmetry, Eulerian approaches dramatically reduce computational costs by bypassing the stochastic Poisson noise inherent to Monte Carlo particle tracking. Crucially, grid-based solvers directly yield smooth, continuous spatial distributions and energy spectra. Among other advantages, this enables for the precise calculation of gamma-ray emissivities without artificial statistical fluctuations.

Hence, there remains a notable gap in the literature for a lightweight, flexible and dedicated finite-volume tool specifically optimized for one-dimensional spherical symmetry.

In this work, we present \texttt{SAETASS} \citep{saetass_software}, a new numerical solver designed to fill this gap by computing the solution of the time-dependent transport equation in one-dimensional spherical geometry. The code employs a conservative finite-volume discretization that correctly incorporates the geometric factors of the operators, ensuring numerical stability and particle conservation. The solver is modular and efficient, allowing it to be easily coupled to various hydrodynamical background profiles. Particular care has been given to the treatment of the origin and the enforcement of spherical regularity conditions. By providing a robust tool for these specific configurations, this solver enables systematic investigations into the interaction between spatial transport and momentum losses in a fully time-dependent framework.

The paper is structured as follows. In Section~\ref{sec:mathematical_numerical}, we introduce the transport equation in spherical symmetry and discuss the physical processes and numerical schemes implemented, including the operator splitting strategy and finite-volume discretization. Section~\ref{sec:tests} presents a set of validation tests designed to assess the accuracy, convergence and reliability for physical application of \texttt{SAETASS}. In Section~\ref{sec:applications}, we illustrate the capabilities of the code through a representative astrophysical application involving a real spherical system. Finally, Section~\ref{sec:conclusions} summarizes our main results and outlines future developments.

\section{The mathematical backbone of \texttt{SAETASS}}
\label{sec:mathematical_numerical}

The numerical solver presented in this work is engineered to solve the time-dependent transport equation in one-dimensional spherical symmetry with high reliability across regimes. The design choices were driven by three goals: to preserve exact conservation properties when they exist, which led to the choice of a finite-volume conservative formulation; to maintain stability and monotonicity in the presence of steep gradients and strong losses, thus using high-order methods; and to provide modularity and flexibility so that operators, splitting strategies and physical modules can be extended or replaced by the user. In what follows we give a complete, self-contained description of the numerical methods: the continuous equations we discretize, the chosen variable transforms, the grid and finite-volume formulation, the operator-splitting philosophy, the detailed algorithms used for hyperbolic and parabolic operators including full discretized expressions, boundary treatment and timestep control. Where alternative approaches are common in the literature we explain why a given choice is appropriate here and what trade-offs it entails.

Specific technical details on the software design, its structure and its implementation can be found in Appendix~\ref{app:software_design}. Furthermore, this section serves as an introduction and outline of the mathematical and numerical foundations of the solver. Beyond this, additional information on the specific computational implementation can be found in the official \texttt{SAETASS} documentation\footnote{Accessible in \href{https://saetass.readthedocs.io/en/latest/}{https://saetass.readthedocs.io/en/latest/}.}.

\subsection{Mathematical and physical introduction}

Equation of transport \eqref{eq:general_transport_equation} is written in a general, vectorized, coordinate-free form. Choosing spherical coordinates and, in particular, imposing spherical symmetry, formally means to consider the following relations:
\begin{gather}
    \label{eq:symmetry_relations}
    \mathbf{w} = w_r\hat{\mathbf{e}}_r,\qquad g(\mathbf{r}) = g(r),
\end{gather}
for any vector field $\mathbf{w}$ and scalar field $g(\mathbf{r})$; where $\hat{\mathbf{e}}_r$ is the unit vector in the direction of $\mathbf{r}$ and $\mathbf{r}=r\hat{\mathbf{e}}_r$ is the position vector. Imposing \eqref{eq:symmetry_relations} to \eqref{eq:general_transport_equation} yields
\begin{equation}
\label{eq:spherical_transport_equation}
    \frac{\partial f}{\partial t} 
    + \frac{1}{r^2}\frac{\partial}{\partial r}\left(r^2 u_\mathrm{w}f\right)
    + \frac{\partial}{\partial p}\left( \dot{p} f \right)
    = \frac{1}{r^2} \frac{\partial}{\partial r}
    \left(r^2 D\frac{\partial f}{\partial r}\right)
    + Q,
\end{equation}
where $f=f(t,r,p)$, $Q=Q(t,r,p)$ and the transport coefficients have analogous dependence: $u_\mathrm{w}=u_\mathrm{w}(t,r,p)$, $\dot{p}=\dot{p}(t,r,p)$ and $D=D(t,r,p)$.

The transport equation as expressed in \eqref{eq:spherical_transport_equation} represents a high-dimensional, non-linear partial differential equation where the interplay between spatial transport, momentum-space energy changes and diffusion can lead to significant numerical stiffness. While the full equation may appear complex due to the coupled nature of the derivatives and the coordinate-dependent coefficients, its underlying structure is more easily understood by decomposing it into several distinct physical operators. 

From a numerical perspective, this decomposition allows for the application of specialized algorithms tailored to the mathematical nature of each term as we will see in Section~\ref{sect:time_discretization_operator_splitting}. We can rewrite equation~\eqref{eq:spherical_transport_equation} as:
\begin{equation}
    \label{eq:operator_decomposition}
    \frac{\partial f}{\partial t} = \mathcal{L}[f] = \mathcal{L}_{\mathrm{adv}}[f] + \mathcal{L}_{\mathrm{diff}}[f] + \mathcal{L}_{\mathrm{loss}}[f] + \mathcal{L}_{\mathrm{src}}[f],
\end{equation}
where the operators are defined and characterized as follows.
\begin{description}
    \item[{Advection Operator ($\mathcal{L}_{\mathrm{adv}}[f]$):}] Defined as
    \begin{equation}
        \label{eq:adv_operator}
        \mathcal{L}_{\mathrm{adv}}[f]=-\frac{1}{r^2}\frac{\partial}{\partial r}(r^2 u_\mathrm{w} f),
    \end{equation}
    this term represents the bulk transport of particles with an advection velocity of $u_\mathrm{w}$, such as that driven by a stellar wind. Mathematically, it is a \textit{hyperbolic} term. In a numerical context, it is prone to producing artificial oscillations or excessive numerical diffusion, necessitating the use of high-resolution upwind schemes or flux limiters to maintain the positivity and monotonicity of the solution.
    
    \item[{Diffusion Operator ($\mathcal{L}_{\mathrm{diff}}[f]$):}] Defined as
    \begin{equation}
        \label{eq:diff_operator}
        \mathcal{L}_{\mathrm{diff}}[f]=\frac{1}{r^2} \frac{\partial}{\partial r} \left(r^2 D\frac{\partial f}{\partial r}\right),
    \end{equation}
    this operator accounts for stochastic scattering processes in the hydrodynamical limit. It is a \textit{parabolic} term, characterized by a smoothing effect on the distribution function. Unlike the hyperbolic terms, diffusion implies a global coupling of the grid cells, which often requires implicit time-stepping to avoid the restrictive stability constraints associated with explicit methods.
    
    \item[{Loss Operator ($\mathcal{L}_{\mathrm{loss}}[f]$):}] Defined as
    \begin{equation}
        \label{eq:loss_operator}
        \mathcal{L}_{\mathrm{loss}}[f]=-\frac{\partial}{\partial p}(\dot{p} f),
    \end{equation}
    this operator describes the variation of the particle energy in momentum space, due to, for example, adiabatic cooling or radiation losses. Although it acts in the $p$-coordinate, it is mathematically \textit{hyperbolic} and analogous to advection. However, because $\dot{p}$ is often negative in astrophysical contexts, it typically represents a flow toward lower momenta, thus leading to the \emph{loss} naming and requiring careful treatment of the inner boundary in momentum space. \texttt{SAETASS} incorporates a suite of predefined loss functions tailored for common astrophysical scenarios, the details of which are provided in Appendix~\ref{app:loss_functions_implemented}.
    
    \item[{Source Operator ($\mathcal{L}_{\mathrm{src}}[f]$):}] Defined simply as
    \begin{equation}
        \label{eq:src_operator}
        \mathcal{L}_{\mathrm{src}}[f]=Q,
    \end{equation}
    this is an algebraic term representing the injection of new particles into the system. It does not involve spatial or momentum derivatives, making it the most straightforward to implement, though its time-dependence may still influence the choice of the integration timestep.
\end{description}

Sometimes, we will refer to these operators by $\mathcal{L}_\alpha$, where $\alpha$ ranges over all the possible subproblems. By isolating these operators, we can make use of the \textit{operator splitting approach}, which is explained in Section~\ref{sect:time_discretization_operator_splitting}. However, before such discussion, some mathematical manipulations can be done to the operators in order to improve the methodology.

\subsection{Generalized variables and operators}
\label{sec:generalized_variable}
In standard astrophysical setups, the momentum coordinate typically spans many orders of magnitude, and so uniform discretization in $p$ either wastes resolution at low momenta or loses accuracy at high momenta. Moreover, severe Courant-Friedrichs-Lewy (CFL) restrictions in momentum-space discretization are often needed when $\dot{p}$ varies across orders of magnitude. In order to handle this efficiently, we adopt the logarithmic variable
\begin{equation*}
    q \equiv \log_{10} p, \qquad p \equiv 10^{q}.
\end{equation*}
Under this change, the momentum derivative transforms accordingly and the loss operator given by \eqref{eq:loss_operator} becomes
\begin{equation}
    \label{eq:loss_operator_transformed}
    \tilde{\mathcal{L}}_{\rm loss}[f] =- \frac{1}{p\ln 10}\frac{\partial}{\partial q}\!\big( \dot p f \big).
\end{equation}

It is easy to see that the subproblems associated to the advection operator defined in \eqref{eq:adv_operator} and the new loss operator as written in \eqref{eq:loss_operator_transformed} can be rewritten in a generalized formulation. In particular, we can define $U=r^2f$ and $V = u_\mathrm{w}$ in the case of advection and rename $U =pf$ and $V = \frac{\dot{p}}{p\ln(10)}$ for the losses case. Then, if we consider a generalized coordinate to denote them, $\xi\in \{r,q\}$, both equations read
\begin{equation}
    \label{eq:general_hyperbolic_eq}
    \frac{\partial U}{\partial t} = -\frac{\partial}{\partial \xi}\left(V\,U\right):=\mathcal{L}_\mathrm{hyp}[U].
\end{equation}
This new formulation by means of a generalized hyperbolic operator allows us to create a general hyperbolic solver that can handle both advection and losses integration.

\subsection{Space and momentum discretization: the finite volume formulation}
\label{sect:space_discretization_finite_volume_formulation}

Once the mathematical structure of the transport equation and its operators has been established, we must choose a numerical framework to discretize the continuous fields. In this work, we adopt the \textit{finite volume method} (FVM) \citep{LeVeque2002} over other common alternatives, such as finite differences (although the finite differences method can be considered a particular case of the finite volume method under some technical conditions \citep{LeVeque2002}) or finite elements method \citep{Brenner2008}. The primary motivation for this choice is the intrinsically conservative nature of the FVM: by discretizing the equation in its integral form, we ensure that the variation of the distribution function within a control volume is exactly balanced by the net fluxes across its boundaries. This property is crucial in astrophysical transport problems where the conservation of the total number of particles (or energy density) is a strict physical requirement. Furthermore, FVM can be robust in the presence of the steep gradients and discontinuities that naturally arise in hyperbolic systems \citep{LeVeque2002}, preventing the appearance of non-physical oscillations that often characterises non-conservative schemes \citep{Chin1975}.

In our implementation, we define a two-dimensional computational domain in the $(r,q)$ plane. The radial grid is one-dimensional and cell-centered with interfaces at $r_{i\pm 1/2}$ and centers at $r_i$, for $i=0,\dots,n_r-1$. The radial intercell areas and cell volumes are given by
\begin{equation*}
    A_{i\pm\frac{1}{2}} = 4\pi r_{i\pm\frac{1}{2}}^2\quad\text{and}\quad V_i = \frac{4\pi}{3}\big( r_{i+1/2}^3 - r_{i-1/2}^3 \big).
\end{equation*}
It is important to note that particular care needs to be taken when $r_0=0$, where we impose $r_{-\frac{1}{2}} =r_0$. The momentum grid in $q$ has interfaces at $q_{j\pm 1/2}$ and centres at $q_j$, $j=0,\dots,n_p-1$, with cell width $\Delta q_j = q_{j+1/2} - q_{j-1/2}$.

The finite-volume formulation then is based on advancing in time the cell-average
\begin{equation}
\label{eq:general_FVM_formulation}
f_{i,j}(t) \equiv \frac{1}{V_i \Delta q_j}\int_{r_{i-1/2}}^{r_{i+1/2}}\int_{q_{j-1/2}}^{q_{j+1/2}} 4\pi r^2f(t,r,q)\, \mathrm{d}q \, \mathrm{d}r
\end{equation}
by computing the fluxes through the intercell faces, thus guaranteeing global conservation of particles across the domain, as we mentioned.

Integrating directly over the equation~\eqref{eq:spherical_transport_equation} does provide a finite volume formulation of the complete problem; however, this is not desirable from the numerical point of view. Instead, it is better to adopt an \emph{operator-splitting} strategy \citep{McLachlan2002, Blanes2024}, which is discussed in Section~\ref{sect:time_discretization_operator_splitting}.

\subsection{Time discretization: the operator-splitting approach}
\label{sect:time_discretization_operator_splitting}

A monolithic numerical routine for the full unsplit formulation obtained after integrating over \eqref{eq:spherical_transport_equation} as given by \eqref{eq:general_FVM_formulation} leads to inefficient and non-specific calculations, often unable to properly account for the different scales and physical phenomena represented by each term of the equation. The full operator in equation~\eqref{eq:spherical_transport_equation} itself illustrates this clearly: it contains advective and loss components (hyperbolic terms), diffusive components (parabolic term) and a explicit source that needs to be carefully treated. In practice, every operator has its own characteristic timescales and numerical requirements, and forcing a unified treatment can introduce stiffness, degrade accuracy and impose stability constraints dictated by the most restrictive part of the system. As a result, the integration of otherwise well-behaving operators becomes unnecessarily overconstrained, reducing both performance and flexibility \citep{LeVeque2002, Toro2009}.

On the contrary, the operator-splitting approach \citep{McLachlan2002, Blanes2024} reduces complexity, allows the use of solvers specifically tailored to each operator and provides a modular structure that is easy to extend or modify. Its main drawback is the introduction of splitting errors, which must be properly quantified and controlled \citep{LeVeque1990}. The full transport operator can be decomposed schematically as was shown in \eqref{eq:operator_decomposition}.  The goal of operator splitting is to approximate the complete operator, $\mathcal{L}$, by a sequence of substeps involving the individual operators, 
$\mathcal{L}_\alpha$, each of which can be treated with a numerical method adapted to its mathematical typology. This way, each operator is advanced in succession and the composition of these substeps approximates the evolution under the full $\mathcal{L}$.

The temporal evolution of the system is described by a discrete sequence of output times $\{t_n\}_{n=0}^{n_t-1}$, over which the numerical solution is advanced from an initial condition $f(t_0,r, p)$ to the final time $t_{n_t}$. We will use the notation $f^n\equiv f(t_n,\cdot ,\cdot )$. Between two consecutive output times $t_n$ and $t_{n+1}$, the solver advances the state by a timestep $\Delta t_n = t_{n+1} - t_n$, applying the sequence of physical operators according to the chosen splitting scheme.

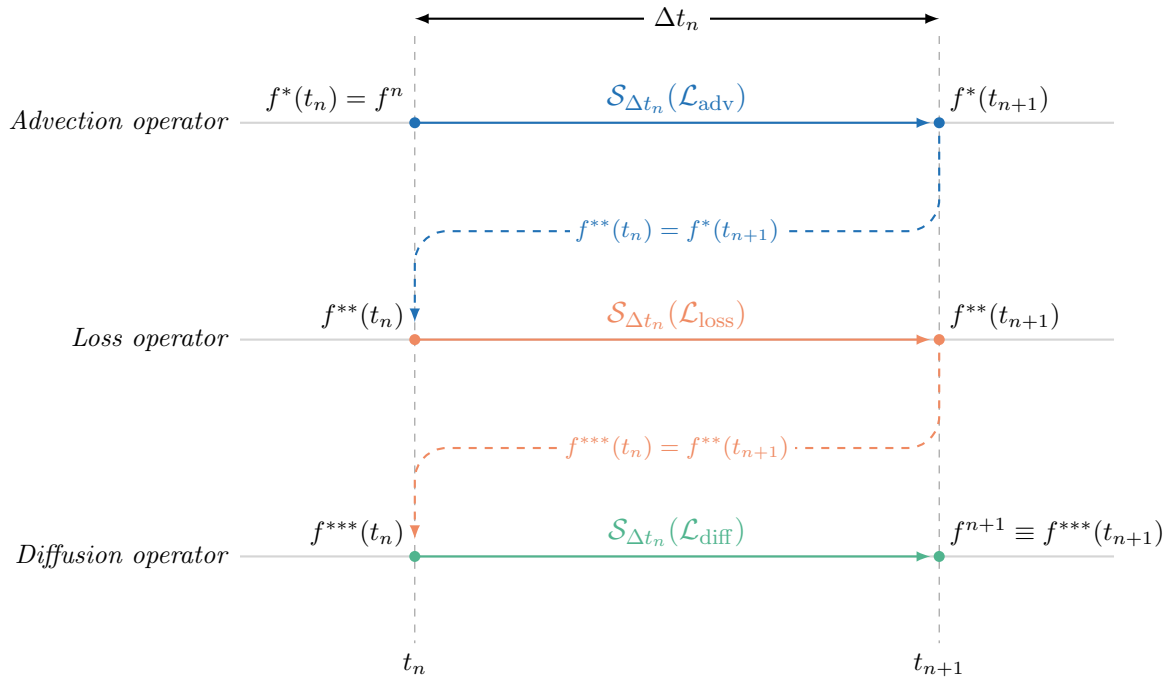
\begin{figure}[ht]
    \centering
\begin{tikzpicture}[scale=1.15,>=latex]

\definecolor{C1}{RGB}{33,113,181}   
\definecolor{C2}{RGB}{239,138,98}   
\definecolor{C3}{RGB}{80,180,142}  
\definecolor{C4}{RGB}{150,150,150}  

\def\tn{2}
\def\tnp{8}

\draw[dashed, C4] (\tn, 1) -- (\tn, -6) node[below, black] {\small $t_n$};
\draw[dashed, C4] (\tnp, 1) -- (\tnp, -6) node[below, black] {\small $t_{n+1}$};

\draw[<->, thick] (\tn, 1.2) -- (\tnp, 1.2) node[midway, fill=white] {\small $\Delta t_n$};

\def\yadv{0}
\draw[thick, C4!40] (0,\yadv) -- (10,\yadv); 
\node[left, black] at (0,\yadv) {\small \textit{Advection operator}};

\draw[->, thick, C1] (\tn,\yadv) -- (\tnp-0.1,\yadv);
\fill[C1] (\tn,\yadv) circle (1.8pt) node[above left, black] {\small $f^*(t_n) = f^n$};
\fill[C1] (\tnp,\yadv) circle (1.8pt) node[above right, black] {\small $f^*(t_{n+1})$};
\node[above, C1] at ({(\tn+\tnp)/2}, \yadv) {$\mathcal{S}_{\Delta t_n}(\mathcal{L}_{\mathrm{adv}})$};

\def\yloss{-2.5}
\draw[->, thick, C1, dashed, rounded corners=12pt] 
    (\tnp, \yadv-0.15) -- (\tnp, \yadv-1.25) -- (\tn, \yadv-1.25) -- (\tn, \yloss+0.2);
\node[fill=white, text=C1, inner sep=2pt] at ({(\tn+\tnp)/2}, \yadv-1.25) 
    {\footnotesize $f^{**}(t_n) = f^*(t_{n+1})$};

\draw[thick, C4!40] (0,\yloss) -- (10,\yloss);
\node[left, black] at (0,\yloss) {\small \textit{Loss operator}};

\draw[->, thick, C2] (\tn,\yloss) -- (\tnp-0.1,\yloss);
\fill[C2] (\tn,\yloss) circle (1.8pt) node[above left, black] {\small $f^{**}(t_n)$};
\fill[C2] (\tnp,\yloss) circle (1.8pt) node[above right, black] {\small $f^{**}(t_{n+1})$};
\node[above, C2] at ({(\tn+\tnp)/2}, \yloss) {$\mathcal{S}_{\Delta t_n}(\mathcal{L}_{\mathrm{loss}})$};

\def\ydiff{-5.0}
\draw[->, thick, C2, dashed, rounded corners=12pt] 
    (\tnp, \yloss-0.15) -- (\tnp, \yloss-1.25) -- (\tn, \yloss-1.25) -- (\tn, \ydiff+0.2);
\node[fill=white, text=C2, inner sep=2pt] at ({(\tn+\tnp)/2}, \yloss-1.25) 
    {\footnotesize $f^{***}(t_n) = f^{**}(t_{n+1})$};

\draw[thick, C4!40] (0,\ydiff) -- (10,\ydiff);
\node[left, black] at (0,\ydiff) {\small \textit{Diffusion operator}};

\draw[->, thick, C3] (\tn,\ydiff) -- (\tnp-0.1,\ydiff);
\fill[C3] (\tn,\ydiff) circle (1.8pt) node[above left, black] {\small $f^{***}(t_n)$};
\fill[C3] (\tnp,\ydiff) circle (1.8pt) node[above right, black] {\small $f^{n+1} \equiv f^{***}(t_{n+1})$};
\node[above, C3] at ({(\tn+\tnp)/2}, \ydiff) {$\mathcal{S}_{\Delta t_n}(\mathcal{L}_{\mathrm{diff}})$};

\end{tikzpicture}
    \caption{Illustration of an example sequential operator-splitting approach to advance the solution function $f$ over a timestep $\Delta t_n =t_{n+1}-t_n$. First, the advection operator solves for $f^*$ starting from $t_n$ until $t_{n+1}$. Then, a dashed transfer assigns the final result $f^*(t_{n+1})$ as the initial condition $f^{**}(t_n)$ for the loss operator, which then solves until $t_{n+1}$. Similarly, the final result $f^{**}(t_{n+1})$ is assigned as the initial condition $f^{***}(t_n)$ for the diffusion operator, yielding the final numerical solution $f^{n+1}\equiv f^{***}(t_{n+1})$. Specifically, this corresponds to a \emph{Lie splitting} \citep{Lie1970, Trotter1959}.}
    \label{fig:operator_splitting_scheme}
\end{figure}

To formalize this, an \textit{evolution operator}, $\mathcal{S}_{\Delta t}(\mathcal{L}_\alpha)$, can be defined as the numerical solution map that advances the distribution function $f$ over a time interval $\Delta t$, this is, for example, advancing the solution from $f^n$ to $f^{n+1}$. In practice, the operator-splitting routine is based on a multi-stage process where the output of one operator serves as the initial condition for the next. For instance, if we consider the particular sequential integration of the advection, loss and diffusion terms, the procedure would follow:
\begin{enumerate}
    \item Solve $\partial_t f^* = \mathcal{L}_{\mathrm{adv}}[f^*]$ over $t \in [t_n, t_{n+1}]$ with $f^*(t_n)=f^n$. This is, calculate $f^*=\mathcal{S}_{\Delta t_n}(\mathcal{L}_\mathrm{adv}[f])$.
    \item Solve $\partial_t f^{**} = \mathcal{L}_{\mathrm{loss}}[f^{**}]$ over the same interval with the initial condition $f^{**}(t_n) = f^*(t_{n+1})$. This is, calculate $f^{**}=\mathcal{S}_{\Delta t_n}(\mathcal{L}_\mathrm{loss}[f^*])$.
    \item Solve analogously $\partial_t f^{***} = \mathcal{L}_{\mathrm{diff}}[f^{***}]$ with $f^{***}(t_n) = f^{**}(t_{n+1})$. So, calculate $f^{***}=\mathcal{S}_{\Delta t_n}(\mathcal{L}_\mathrm{diff}[f^{**}])$.
    \item Finally, $f^{n+1}\equiv f^{***}(t_{n+1})$ is obtained.
\end{enumerate}
This sequential composition can be written more compactly using the operator notation as $f^{n+1} = \mathcal{S}_{\Delta t_n}(\mathcal{L}_{\mathrm{diff}}) \circ \mathcal{S}_{\Delta t_n}(\mathcal{L}_{\mathrm{loss}}) \circ \mathcal{S}_{\Delta t_n}(\mathcal{L}_{\mathrm{adv}})[f^n]$. A schematic representation of this process is illustrated in Figure \ref{fig:operator_splitting_scheme}.

Depending on the order in which these operators are applied and how the time step is subdivided, different splitting schemes with varying degrees of accuracy can be constructed. In our solver, we have implemented the two most widely used strategies, which can be then chosen by the user, as well as the order of operator application, according to the problem being simulated. 

First, we implemented a well-known first-order approach, the \textit{Lie splitting} \citep{Lie1970} (sometimes also called \textit{Lie-Trotter splitting} \citep{Trotter1959}), which is exactly the algorithm we have introduced. An example of a specific ordering, in a setup in which all operators are considered, is
\begin{equation*}
    f^{n+1} = \mathcal{S}_{\Delta t}(\mathcal{L}_{\mathrm{adv}}) \circ \mathcal{S}_{\Delta t}(\mathcal{L}_{\mathrm{diff}}) \circ \mathcal{S}_{\Delta t}(\mathcal{L}_{\mathrm{loss}}) \circ \mathcal{S}_{\Delta t}(\mathcal{L}_{\mathrm{src}})[f^{n}].
\end{equation*}

On the other hand, a second-order alternative is the \textit{Strang splitting} \citep{Strang1968}, which symmetrizes the composition and reduces the splitting error. It is defined by
\begin{equation*}
    f^{n+1} = \mathcal{S}_{\frac{\Delta t}{2}}(\mathcal{L}_{\mathrm{diff}}) \circ \mathcal{S}_{\frac{\Delta t}{2}}(\mathcal{L}_{\mathrm{loss}}) \circ \mathcal{S}_{\Delta t}(\mathcal{L}_{\mathrm{src}}) \circ \mathcal{S}_{\frac{\Delta t}{2}}(\mathcal{L}_{\mathrm{loss}}) \circ \mathcal{S}_{\frac{\Delta t}{2}}(\mathcal{L}_{\mathrm{diff}})  [f^{n}],
\end{equation*}
in a setup in which, for example, we are not interested in advection.

Lie splitting is light and easy to implement, whereas Strang splitting achieves higher accuracy at the cost of additional substeps. Specific splitting strategies designed by the user are also supported.

\subsection{Numerical treatment of the physical operators}
\label{sec:numerical_operators}

Having established the finite volume framework and the operator-splitting strategy, we now detail the specific numerical algorithms applied to each isolated subproblem. The distinct mathematical nature of each operator strictly dictates its optimal numerical treatment. Hyperbolic terms, representing radial advection and continuous momentum losses, are prone to developing steep gradients and require robust shock-capturing techniques. This is discussed in Section~\ref{sec:hyperbolic_MUSCL_Hancock}. On the other hand, the parabolic diffusion operator, which would otherwise impose severe explicit timestep restrictions due to global cell coupling, is integrated using an implicit algorithm discussed in Section~\ref{sect:diffusion_operator_implementation}. Finally, localized source injections are treated through explicit algebraic updates. We show this in Section~\ref{sec:source_operator_implementation}.

Across all implementations, strict adherence to the finite-volume conservation laws and the proper geometric treatment of the spherical domain remain the guiding principles.

\subsubsection{Hyperbolic operators: MUSCL-Hancock scheme for radial advection and momentum losses}
\label{sec:hyperbolic_MUSCL_Hancock}

Radial advection and momentum-loss operators are generalized following the expressions given in Section~\ref{sec:generalized_variable}. Thus, it is natural to adopt a scheme tailored to this conservative hyperbolic formulation. The standard approach to this numerical problem is Godunov's method \citep{Godunov1959}, whose general philosophy can be understood by integrating the equation \eqref{eq:general_hyperbolic_eq} over a computational cell $[\xi_{k-1/2}, \xi_{k+1/2}]$:
\begin{equation*}
    \frac{\partial U_k}{\partial t}
    = \frac{1}{\Delta \xi_k} \int_{\xi_{k-1/2}}^{\xi_{k+1/2}}
      \left(-\frac{\partial}{\partial \xi}(V\,U)\right)\, \mathrm{d}\xi
    \;\implies\;
    \frac{\mathrm{d} U_k}{\mathrm{d} t}
    = -\frac{1}{\Delta \xi_k}\bigg(F(t,U)\big|_{\xi_{k+1/2}} - F(t,U)\big|_{\xi_{k-1/2}}\bigg),
\end{equation*}
where $U_k$ is the cell-averaged value and $F(t,U)|_{\xi_{k\pm1/2}}$ represents the physical flux crossing the interface at $\xi_{k\pm1/2}$ at time $t$. In our case, $F(t,U) = V(t,\xi)U(t,\xi)$.

A fundamental challenge arises here: while the FVM evolves the cell-averaged quantities $U_k$, the fluxes must be evaluated at the interfaces $\xi_{k \pm 1/2}$, where the solution is not explicitly known. To resolve this, one must \textit{reconstruct} the pointwise values of $U$ at the interfaces from the available cell averages. Since this reconstruction is performed independently for each cell, the process naturally yields two potentially different values at any given interface $\xi_{k+1/2}$: a left state $U^L_{k+1/2}$ (extrapolated from cell $k$) and a right state $U^R_{k+1/2}$ (extrapolated from cell $k+1$). 

The interaction between these two states is what defines the numerical flux $\mathcal{F}(U^L, U^R)$, which effectively solves a local \emph{Riemann problem} \citep{Toro2009} at the boundary. The general discrete update for Godunov-type methods is therefore expressed as
\begin{equation}
    \label{eq:time_advance_from_numerical_flux}
    U_k^{n} = U_k^{n-1} + \frac{\Delta t}{\Delta \xi_k} \left( \mathcal{F}^{n-1}_{k-1/2} - \mathcal{F}^{n-1}_{k+1/2} \right).
\end{equation}

The different versions of Godunov's method essentially differ in the type of reconstruction performed and the specific solver used for computing $\mathcal{F}$. In the original first-order Godunov scheme \citep{Godunov1959}, $U(\xi)$ is assumed to be piecewise constant within each cell, meaning $U^L_{k+1/2} = U_k$ and $U^R_{k+1/2} = U_{k+1}$.

\begin{figure}[ht]
    \centering
\begin{tikzpicture}[scale=1, >=latex]

\definecolor{C1}{RGB}{33,113,181}   
\definecolor{C2}{RGB}{239,138,98}   
\definecolor{C3}{RGB}{80,180,142}  
\definecolor{C4}{RGB}{150,150,150}  

\def\xfA{0}     
\def\xfB{2}     
\def\xfC{4.5}   
\def\xfD{9}     
\def\xfE{11.5}  
\def\xfF{13.5}  

\def\xcA{1}     
\def\xcB{3}     
\def\xcC{6.5}   
\def\xcD{10.5}  
\def\xcE{12.5}  

\def\uA{0.8}    
\def\uB{1.7}    
\def\uC{3.2}    
\def\uD{2.0}    
\def\uE{0.8}    

\def\sA{0.15}
\def\sB{0.4}
\def\sC{0.25}
\def\sD{-0.3}
\def\sE{-0.15}

\draw[thick, ->] (-0.5, 0) -- (14.2, 0) node[right, black] {$\xi$};

\draw[dashed, C4] (\xfA, 0) -- (\xfA, 2);
\draw[dashed, C4] (\xfB, 0) -- (\xfB, 3);
\draw[dashed, C4] (\xfC, 0) -- (\xfC, 4.5);
\draw[dashed, C4] (\xfD, 0) -- (\xfD, 4.5);
\draw[dashed, C4] (\xfE, 0) -- (\xfE, 3.5);
\draw[dashed, C4] (\xfF, 0) -- (\xfF, 2);

\node[below, black] at (\xfA, -0.1) {\small $\xi_{k-5/2}$};
\node[below, black] at (\xfB, -0.1) {\small $\xi_{k-3/2}$};
\node[below, black] at (\xfC, -0.1) {\small $\xi_{k-1/2}$};
\node[below, black] at (\xfD, -0.1) {\small $\xi_{k+1/2}$};
\node[below, black] at (\xfE, -0.1) {\small $\xi_{k+3/2}$};
\node[below, black] at (\xfF, -0.1) {\small $\xi_{k+5/2}$};

\node[below, black] at (\xcA, -0.1) {\small $\xi_{k-2}$};
\node[below, black] at (\xcB, -0.1) {\small $\xi_{k-1}$};
\node[below, black] at (\xcC, -0.1) {\small $\xi_k$};
\node[below, black] at (\xcD, -0.1) {\small $\xi_{k+1}$};
\node[below, black] at (\xcE, -0.1) {\small $\xi_{k+2}$};

\draw[dotted, C4!80!black] (\xcA, 0) -- (\xcA, \uA);
\draw[dotted, C4!80!black] (\xcB, 0) -- (\xcB, \uB);
\draw[dotted, C4!80!black] (\xcC, 0) -- (\xcC, \uC);
\draw[dotted, C4!80!black] (\xcD, 0) -- (\xcD, \uD);
\draw[dotted, C4!80!black] (\xcE, 0) -- (\xcE, \uE);

\draw[<->, C4!80!black] (\xfC, 0.5) -- (\xcC, 0.5) node[midway, above, inner sep=1pt] {\footnotesize $\Delta \xi_k^L$};
\draw[<->, C4!80!black] (\xcC, 0.5) -- (\xfD, 0.5) node[midway, above, inner sep=1pt] {\footnotesize $\Delta \xi_k^R$};

\draw[C1!40, thick, dashed] (\xfA, \uA) -- (\xfB, \uA);
\draw[C1, thick] (\xfA, {\uA - \sA*(\xcA - \xfA)}) -- (\xfB, {\uA + \sA*(\xfB - \xcA)});
\fill[C1] (\xcA, \uA) circle (1.8pt) node[above, C1] {$U_{k-2}$};
\fill[C2] (\xfB, {\uA + \sA*(\xfB - \xcA)}) circle (2pt) node[below right, C2, inner sep=1pt] {$U_{k-3/2}^L$};

\draw[C1!40, thick, dashed] (\xfB, \uB) -- (\xfC, \uB);
\draw[C1, thick] (\xfB, {\uB - \sB*(\xcB - \xfB)}) -- (\xfC, {\uB + \sB*(\xfC - \xcB)});
\fill[C1] (\xcB, \uB) circle (1.8pt) node[above, C1] {$U_{k-1}$};
\fill[C3] (\xfB, {\uB - \sB*(\xcB - \xfB)}) circle (2pt) node[above left, C3, inner sep=1pt] {$U_{k-3/2}^R$};
\fill[C2] (\xfC, {\uB + \sB*(\xfC - \xcB)}) circle (2pt) node[below right, C2, inner sep=1pt] {$U_{k-1/2}^L$};

\draw[C1!40, thick, dashed] (\xfC, \uC) -- (\xfD, \uC);
\draw[C1, thick] (\xfC, {\uC - \sC*(\xcC - \xfC)}) -- (\xfD, {\uC + \sC*(\xfD - \xcC)});
\fill[C1] (\xcC, \uC) circle (1.8pt) node[above, C1] {$U_k$};
\fill[C3] (\xfC, {\uC - \sC*(\xcC - \xfC)}) circle (2pt) node[above left, C3, inner sep=1pt] {$U_{k-1/2}^R$};
\fill[C2] (\xfD, {\uC + \sC*(\xfD - \xcC)}) circle (2pt) node[above left, C2, inner sep=1pt] {$U_{k+1/2}^L$};

\coordinate (Cang) at (\xcC+1.2, \uC);
\coordinate (Bang) at (\xcC, \uC);
\coordinate (Aang) at (\xcC+1.2, {\uC + \sC*1.2});
\draw[C1!60, thin] (\xcC, \uC) -- (\xcC+1.2, \uC);
\pic[
    draw=C1, 
    ->, 
    "\textcolor{C1}{\footnotesize $\sigma_k$}", 
    angle eccentricity=1.4, 
    angle radius=13mm
] {angle = Cang--Bang--Aang};

\draw[C1!40, thick, dashed] (\xfD, \uD) -- (\xfE, \uD);
\draw[C1, thick] (\xfD, {\uD - \sD*(\xcD - \xfD)}) -- (\xfE, {\uD + \sD*(\xfE - \xcD)});
\fill[C1] (\xcD, \uD) circle (1.8pt) node[above, C1] {$U_{k+1}$};
\fill[C3] (\xfD, {\uD - \sD*(\xcD - \xfD)}) circle (2pt) node[above right, C3, inner sep=1pt] {$U_{k+1/2}^R$};
\fill[C2] (\xfE, {\uD + \sD*(\xfE - \xcD)}) circle (2pt) node[above right, C2, inner sep=1pt] {$U_{k+3/2}^L$};

\draw[C1!40, thick, dashed] (\xfE, \uE) -- (\xfF, \uE);
\draw[C1, thick] (\xfE, {\uE - \sE*(\xcE - \xfE)}) -- (\xfF, {\uE + \sE*(\xfF - \xcE)});
\fill[C1] (\xcE, \uE) circle (1.8pt) node[above, C1] {$U_{k+2}$};
\fill[C3] (\xfE, {\uE - \sE*(\xcE - \xfE)}) circle (2pt) node[above left, C3, inner sep=1pt] {$U_{k+3/2}^R$};

\end{tikzpicture}
    \caption{Schematic representation of the MUSCL-Hancock spatial discretization over a five-cell non-uniform mesh in the $\xi$-domain, ranging from cell $k-2$ to $k+2$. The dark blue nodes ($U_k$) represent the cell-averaged values, while solid blue lines denote the piecewise linear reconstruction governed by the local slopes $\sigma_k$. The mesh non-uniformity is highlighted in cell $k$ by the asymmetric distances $\Delta \xi_k^L$ and $\Delta \xi_k^R$. At each interface $\xi_{k \pm 1/2}$, the reconstruction yields a discontinuity defined by the left-extrapolated state, $U^L$ (orange), and the right-extrapolated state, $U^R$ (green), which serve as inputs for the Riemann solver (in our case, simple upwinding).}
    \label{fig:FVM_slope_discretization}
\end{figure}
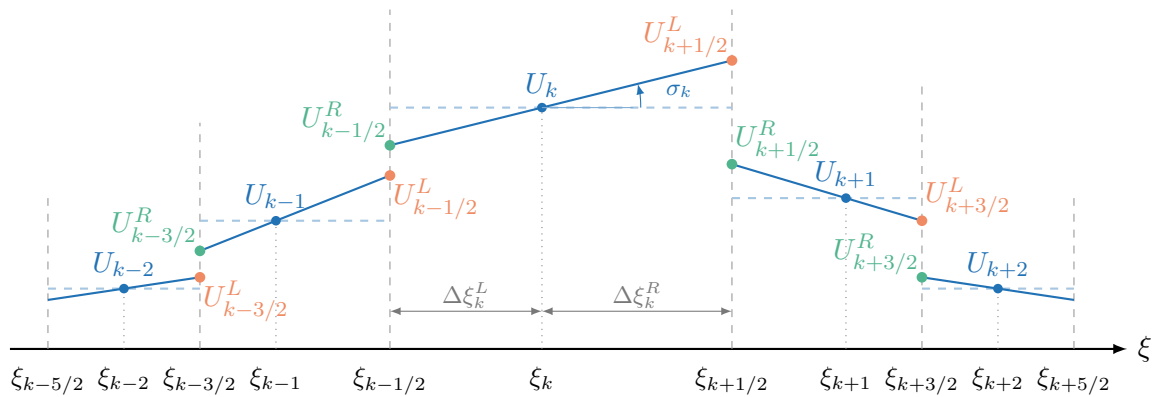

A natural extension of the Godunov method that increases the order of accuracy while retaining its conservative properties is the \emph{MUSCL-Hancock scheme} \citep{vanLeer1977, vanLeer1979, vanLeer1984}. In this approach, the solution within each cell is no longer approximated as piecewise constant, but as a piecewise linear function. This is constructed from the cell-averaged values $U_k$ and limited slopes $\sigma_k$ that prevent the introduction of new, non-physical extrema. Specifically, for each cell, the linear reconstruction reads
\begin{equation}
    \label{eq:slope_reconstruction}
    U_k(\xi) = U_k + \sigma_k\,(\xi-\xi_k).
\end{equation}
A graphical representation of this specific reconstruction strategy is presented in Figure~\ref{fig:FVM_slope_discretization}.

Because the generalized velocity field $V(t,\xi)$ may present strong spatial gradients or temporal dependencies, our implementation of the MUSCL-Hancock method evaluates both the states and the advective fluxes dynamically. To properly handle non-uniform meshes, we denote the distance from the cell center to its left and right interfaces as $\Delta \xi_k^L = \xi_k - \xi_{k-1/2}$ and $\Delta \xi_k^R = \xi_{k+1/2} - \xi_k$, respectively. The integration from $t_{n-1}$ to $t_n = t_{n-1} + \Delta t_{n-1}$ proceeds through the following sequence of operations:

\begin{enumerate}
    \item \textbf{Velocity field evaluation:} First, the time-dependent generalized velocities are evaluated at the cell centers, $V_k^{n-1} = V(t_{n-1}, \xi_k)$. These are then interpolated to the cell interfaces via distance-weighted linear interpolation to obtain $V_{k\pm1/2}^{n-1}$.
    
    \item \textbf{Slope computation:} The spatial derivatives of the conserved variable, $U_k^{n-1}$, are computed using neighbouring cells. To ensure total variation diminishing (TVD) properties and avoid spurious oscillations near discontinuities, these derivatives are restricted using a non-linear slope limiter \citep{Harten1983} (such as minmod, van Leer or monotonized central \citep{LeVeque2002, Toro2009}), yielding the limited slopes $\sigma_k$.
    
    \item \textbf{Spatial reconstruction:} The limited slopes are used to extrapolate the cell-centered values to the internal interfaces. For any given interface at $\xi_{k+1/2}$, this yields two boundary states: one extrapolated from the left cell ($U^L$) and one from the right cell ($U^R$)\footnote{Notation might result heavy. Note that $L$ and $R$ superscripts are meant to be understood \emph{with respect to their respective $k$-index subscript}. This is, for instance, $\Delta \xi_k^R$ is the distance from $\xi_k$ to \emph{its right} closest intercell; whereas $U_{k+1/2}^R$ denotes the extrapolated value at the \emph{right side} of the $(k+\frac{1}{2})$-th intercell. Figure~\ref{fig:FVM_slope_discretization} may result clarifying in this regard.}:
    \begin{equation*}
        U^{L}_{k+1/2} = U_k^{n-1} + \sigma_k \Delta \xi_k^R, \qquad
        U^{R}_{k+1/2} = U_{k+1}^{n-1} - \sigma_{k+1} \Delta \xi_{k+1}^L.
    \end{equation*}
    
    \item \textbf{Predictor step:} An intermediate half-timestep ($\Delta t_{n-1} / 2$) evolution is performed to advance the boundary states to $t_{n-1/2}$, accounting for the local flux gradients. In our framework, this is achieved by computing limited slopes of the advective flux, $\sigma_k^F$, derived from the cell-centered fluxes $F_k = V_k^{n-1} U_k^{n-1}$. The boundary states are thus predicted as:
    \begin{equation*}
        U^{\star, L}_{k+1/2} = U^L_{k+1/2} - \frac{\Delta t_{n-1}}{2} \sigma_k^F, \qquad
        U^{\star, R}_{k+1/2} = U^R_{k+1/2} - \frac{\Delta t_{n-1}}{2} \sigma_{k+1}^F.
    \end{equation*}
    
    \item \textbf{Riemann solver (upwinding):} The numerical flux at each interface is determined by solving a local Riemann problem. Given the advective nature of the generalized operator, we apply a standard upwind scheme based on the sign of the interpolated interface velocity:
    \begin{equation*}
        \mathcal{F}_{k+1/2}^{n-1} = 
        \begin{cases}
        V_{k+1/2}^{n-1}\, U^{\star,L}_{k+1/2}, & \text{if }V_{k+1/2}^{n-1} \ge 0,\\
        V_{k+1/2}^{n-1}\, U^{\star,R}_{k+1/2}, & \text{if }V_{k+1/2}^{n-1} < 0.
        \end{cases}
    \end{equation*}
    
    \item \textbf{Corrector step:} Before the final update, physical boundary conditions are applied by overwriting the numerical fluxes at the domain limits ($\mathcal{F}_{-1/2}$ and $\mathcal{F}_{n_\xi-1/2}$) according to the specific problem geometry and user choice. Finally, the cell averages are advanced to the next time step using the conservative Godunov's formula given by the equation~\eqref{eq:time_advance_from_numerical_flux}.
\end{enumerate}

The MUSCL-Hancock scheme is, hence, a second-order extension of Godunov's method, achieving second-order accuracy in both space and time for smooth solutions through a spatial MUSCL reconstruction combined with a half-step predictor in time \citep{vanLeer1977, vanLeer1979, vanLeer1984, Toro2009}. For one-dimensional homogeneous scalar conservation laws, combining this reconstruction with appropriate slope limiters and adhering to a strict CFL condition yields a TVD finite volume scheme that retains the conservative and shock-capturing properties of Godunov-type methods \citep{Harten1983}. However, this rigorous TVD property does not theoretically extend to nonlinear systems of equations or problems with discontinuous transport coefficients. Consequently, the standard MUSCL-Hancock scheme is not guaranteed to be strictly positivity-preserving for physical variables in such complex, but typical, configurations \citep{Zhang2010}. To prevent the introduction of unphysical negative values caused by reconstruction overshoots at discontinuities, practical implementations must incorporate dedicated positivity-preserving limiters or apply small numerical floors. Details regarding the temporal stability limits (CFL condition) and dynamic timestepping are discussed in Section~\ref{sec:substep_calculation}.

\subsubsection{Diffusive operator: a finite volume Crank-Nicolson implementation}
\label{sect:diffusion_operator_implementation}

The diffusive operator accounts for stochastic scattering processes in the hydrodynamical limit of the transport equation. To ensure global conservation and numerical stability, it is treated using a fully implicit Crank-Nicolson discretization \citep{Crank1947} within our conservative finite volume framework, specifically adapted to the spherical geometry of the problem. Considering the generalized temporal and momentum dependence of the diffusion coefficient, the governing equation for this subproblem, given by the operator defined in \eqref{eq:diff_operator}, reads
\begin{equation}
    \label{eq:diffusion_equation}
    \frac{\partial f}{\partial t}
    = \frac{1}{r^2} \frac{\partial}{\partial r}
    \left(r^2 D(t, r, p)\,\frac{\partial f}{\partial r}\right).
\end{equation}

While standard finite difference implementations of the Crank-Nicolson scheme evaluate the second spatial derivative at grid nodes \citep{LeVeque2007}, our approach integrates the equation~\eqref{eq:diffusion_equation} over a radial control volume $[r_{i-1/2},r_{i+1/2}]$. This finite volume formulation tracks the cell-averaged values $f_i$ and explicitly evaluates the fluxes crossing the intercell boundaries. The spatial semi-discretization takes the form
\begin{equation}
    \label{eq:discrete_diffusion}
    \frac{\mathrm{d} f_i}{\mathrm{d} t} = \frac{1}{V_i} \left[ G_{i+1/2}\left(\frac{f_{i+1}-f_i}{\Delta r_{i+1/2}}\right) - G_{i-1/2}\left(\frac{f_i-f_{i-1}}{\Delta r_{i-1/2}} \right)\right],
\end{equation}
where $V_i$ is the cell volume, $\Delta r_{i\pm1/2}$ are the distances between neighbouring cell centers, and $G_{i\pm1/2} = r_{i\pm1/2}^2\,D_{i\pm1/2}$ represent the geometric conductances through the faces. 

A critical detail in this discretization is the evaluation of the diffusion coefficient at the cell interfaces, $D_{i\pm1/2}$. In astrophysical environments, $D(t,r,p)$ can exhibit steep gradients or even physical discontinuities across different regions. Using a simple arithmetic mean for the interface diffusion would smear these gradients and violate flux conservation. Instead, our solver computes the face values using a distance-weighted harmonic mean,
\begin{equation*}
    D_{i+1/2} = \frac{\Delta r_i^R + \Delta r_{i+1}^L}{\frac{\Delta r_i^R}{D_i} + \frac{\Delta r_{i+1}^L}{D_{i+1}}},
\end{equation*}
where $\Delta r_i^R$ and $\Delta r_{i+1}^L$ are the distances from the $i$-th and $(i+1)$-th cell center to the $(i+\frac{1}{2})$-th face, respectively. This formulation rigorously guarantees the continuity of the diffusive flux across cell interfaces, regardless of the underlying grid spacing or discontinuities in $D$.

To advance the system in time, we apply the Crank-Nicolson method, which averages the fluxes at times $t_n$ and $t_{n+1}$. Substituting this into the equation~\eqref{eq:discrete_diffusion} yields a coupled linear system for each momentum slice,
\begin{equation}
    \label{eq:crank_nicolson}
    \left(\frac{2V_i}{\Delta t_n} + A_i\right) f_i^{n+1} - B_{i,i-1} f_{i-1}^{n+1} - B_{i,i+1} f_{i+1}^{n+1} = \left(\frac{2V_i}{\Delta t_n} - A_i\right) f_i^{n} + B_{i,i-1} f_{i-1}^{n} + B_{i,i+1} f_{i+1}^{n},
\end{equation}
where the coefficients $A_i$ and $B_{i, j}$ absorb the geometric conductances $G_{i\pm1/2}$. This results in a tridiagonal matrix system. Because the diffusion operator does not couple different momentum slices, the equation is solved independently for each $p$ (or, analogously, for each $q$). Our solver achieves high computational efficiency by deploying a vectorized, batched Thomas algorithm \citep{Thomas1949} that solves all momentum slices simultaneously without requiring a sequential loop over $p$.

Regarding boundary conditions, regularity at the origin ($r=0$) is inherently enforced in the spherical finite volume formulation by setting the inner conductance to zero ($G_{-1/2}=0$), naturally imposing a zero-flux symmetry condition. At the outer boundary ($r=r_\mathrm{end}$), the solver dynamically accommodates user-specified conditions, including Dirichlet ($f = f_\mathrm{end}(t)$), Neumann (zero gradient) or asymptotic outflow boundaries.

Finally, a brief note on temporal stability is needed. Analytically, the Crank-Nicolson method is very well known to be unconditionally stable for arbitrarily large timesteps for linear dissipative problems \citep{LeVeque2007}. However, physically strongly damped modes are not efficiently suppressed by the scheme. In practical applications involving strong spatial heterogeneity or sharp gradients, this property can lead to slowly decaying high-frequency numerical modes that may manifest as spurious oscillations when excessively large timesteps are used \cite{Britz2003}. Therefore, although the implicit treatment of diffusion removes the severe timestep restrictions associated with explicit schemes, adequate temporal resolution may still be required to ensure both the physical accuracy and the convergence of the solution.

\subsubsection{Source operator}
\label{sec:source_operator_implementation}

The treatment of source terms in the operator-splitting framework of the finite volume scheme leads to the integration of the evolution equation
\begin{equation*}
    \frac{\partial f}{\partial t} = Q(t, r, p).
\end{equation*}

This source term can be integrated using any time-integration method, depending on the specific properties of $Q$ in every setup. Typically, it takes the form of a simple or exponential cutoff power law. Solving explicitly over a time interval using the simple Euler forward scheme has proven to be sufficiently stable for the use cases of \texttt{SAETASS} at its first version, so the time evolution implemented in the solver reads
\begin{equation*}
    f^{n+1} = f^n + \Delta t_n\, Q^n.
\end{equation*}
However, future versions of \texttt{SAETASS} will include more sophisticated time evolution integrators and the modular structure of the code allows any user to currently introduce their own integration strategies.

Another approach for the treatment of this term is to include $Q$ (or parts of $Q$) with every other operator and, hence, solve the complete non-homogeneous subproblems associated with each of the operators defined in \eqref{eq:adv_operator}, \eqref{eq:diff_operator} and \eqref{eq:loss_operator}. This is in fact another standard technique \citep{Toro2009} and will also be included in future versions of the software.

\subsection{Time stepping, timestep control and splitting timestep strategy}
\label{sec:substep_calculation}

\begin{figure}[ht!]
    \centering
\begin{tikzpicture}[scale=1.15,>=latex]

\definecolor{C1}{RGB}{33,113,181}   
\definecolor{C2}{RGB}{239,138,98}   
\definecolor{C3}{RGB}{80,180,142}  
\definecolor{C4}{RGB}{204,121,167}  

\foreach \y/\label in {0/{Global time grid}, -1.5/{Splitting refinement}, -3/{User substepping}, -4.5/{CFL refinement}} {
  \draw[thick] (0,\y) -- (10,\y) node[right,black] {\small \textit{\label}};
}

\foreach \x in {0,1,2,3,4,5} {
  \draw[C1,thick] (2*\x,0.1) -- (2*\x,-0.1);
  \fill[C1] (2*\x,0) circle (1.8pt);
  \node[above,C1] at (2*\x,0.2) {\scriptsize $t_{\x}$};
}

\foreach \x in {0,2,4,6,8,10} {
  \draw[C1,thick] (\x,-1.4) -- (\x,-1.6);
  \fill[C1] (\x,-1.5) circle (1.5pt);
  \ifnum\x<10
    \pgfmathsetmacro{\xmid}{\x+1}
    \draw[C2,thick] (\xmid,-1.4) -- (\xmid,-1.6);
    \fill[C2] (\xmid,-1.5) circle (1.5pt);
  \fi
}

\foreach \xstart in {0,1,2,3,4,5,6,7,8,9} {
  \pgfmathsetmacro{\xnext}{\xstart+1}
  \foreach \k in {0,1,2,3} {
    \pgfmathsetmacro{\xsub}{\xstart + 0.3333*\k}
    \draw[C3,thick] (\xsub,-2.9) -- (\xsub,-3.1);
    \fill[C3] (\xsub,-3) circle (1.3pt);
  }
}
\foreach \x in {0,2,4,6,8,10} {
  \draw[C1,thick] (\x,-2.9) -- (\x,-3.1);
  \fill[C1] (\x,-3) circle (1.3pt);
}
\foreach \x in {1,3,5,7,9} {
  \draw[C2,thick] (\x,-2.9) -- (\x,-3.1);
  \fill[C2] (\x,-3) circle (1.3pt);
}

\foreach \x in {0,1,...,10} {
  \draw[C1,thick] (\x,-4.4) -- (\x,-4.6);
  \fill[C1] (\x,-4.5) circle (1pt);
}
\foreach \x in {1,3,5,7,9} {
  \draw[C2,thick] (\x,-4.4) -- (\x,-4.6);
  \fill[C2] (\x,-4.5) circle (1pt);
}
\foreach \xstart in {0,1,2,3,4,5,6,7,8,9} {
  \foreach \k in {1,2} {
    \pgfmathsetmacro{\xsub}{\xstart + \k/3} 
    \draw[C3,thick] (\xsub,-4.4) -- (\xsub,-4.6);
    \fill[C3] (\xsub,-4.5) circle (1pt);
  }
}

\foreach \xbase in {0.3333,1.6666,3.0,4.3333,5.6666,7.0,8.3333} {
  \foreach \k in {1,2,3} {
    \pgfmathsetmacro{\xCFL}{\xbase + 0.08*\k} 
    \draw[C4,thick] (\xCFL,-4.4) -- (\xCFL,-4.6);
    \fill[C4] (\xCFL,-4.5) circle (1pt);
  }
}

\foreach \x in {0.5} {
  \draw[->,thick,Blue!70] (\x,-0.2) -- (\x,-1.3);
  \draw[->,thick,Blue!70] (\x,-1.7) -- (\x,-2.8);
  \draw[->,thick,Blue!70] (\x,-3.2) -- (\x,-4.3);
}
\node[Blue!70,right] at (0.6,-2.3) {\footnotesize \textit{Refinement flow}};


\end{tikzpicture}
    \caption{Schematic representation of the temporal refinement hierarchy in the solver. The topmost line (blue ticks) shows the global time grid with coarse timesteps. The second line (orange ticks) illustrates the refinement due to splitting, assuming the operator is included in a Strang splitting scheme and is not the central operator. The third line (green ticks) represents additional user-specified substepping, here chosen as $n_\mathrm{sub} = 3$. Finally, the bottom line (pink ticks) depicts local CFL-driven refinements, applied only where stability constraints require smaller substeps.}
    \label{fig:timestep_refinement_scheme}
\end{figure}
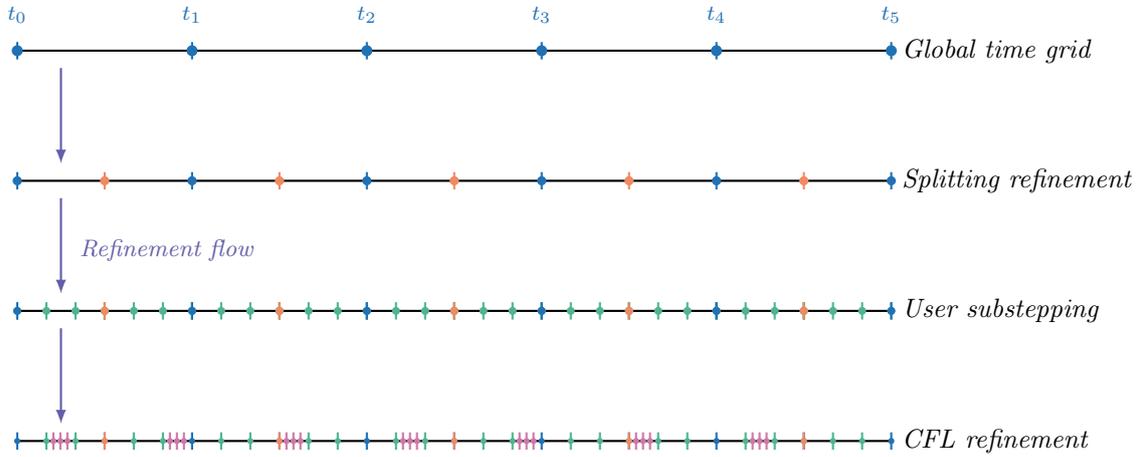

As it was already mentioned, the temporal discretization of our numerical scheme is based on the discrete sequence of output times $\{t_n\}_{n=0}^{n_t-1}$, with $\Delta t_n = t_{n+1} - t_n$. In practice, the nominal timestep $\Delta t_n$, provided by the user during the global temporal grid definition, is refined and adjusted through several hierarchical mechanisms to ensure both accuracy and stability. Such refinement pipeline is graphically represented in Figure~\ref{fig:timestep_refinement_scheme}. These mechanisms can be classified into three levels:

\begin{enumerate}
    \item \textbf{Splitting-induced refinements:} Depending on the chosen operator-splitting scheme, certain operators are evolved over fractional timesteps. Thus, in a fisrt step, the original internal temporal grid, $\{t_n\}_{n=0}^{n_t-1}$, is refined accordingly for each operator. This way, the solver calculates $\Delta t_\mathrm{split}$ reflecting the specific requirements of the chosen splitting method.
    
    \item \textbf{User-prescribed substepping:} Beyond the refinement imposed by the splitting scheme, the user may specify an additional substepping factor $n_\mathrm{sub}$ for each operator. This allows the temporal resolution of selected processes to be enhanced, for instance when certain source or loss terms evolve on shorter timescales than the global dynamics. For an operator with substepping $n_\mathrm{sub}$, its effective internal timestep becomes
    \begin{equation*}
        \Delta t_\mathrm{sub} = \frac{\Delta t_\mathrm{split}}{n_\mathrm{sub}}.
    \end{equation*}
    Each operator then evolves its associated quantity through a sequence of $n_\mathrm{sub}$ substeps, advancing by $\Delta t_\mathrm{sub}$ at each iteration.
    
    \item \textbf{Stability-driven internal control:} Even after the above refinements, some operators may require further internal timestep adjustments to maintain numerical stability. In particular, the hyperbolic transport operators are subject to the standard CFL condition \citep{LeVeque2002},
    \begin{equation*}
        \Delta t_\mathrm{CFL} = C_\mathrm{CFL}\, \frac{\min(\Delta \xi)}{\max(|V|)},
    \end{equation*}
    where $C_\mathrm{CFL}<1$ is the chosen Courant number. At the beginning of each temporal advancement, the operator computes the local $\Delta t_\mathrm{CFL}$ and compares it with the incoming timestep \(\Delta t_\mathrm{sub}\). If the stability condition demands a smaller step, the operator performs multiple sub-iterations using \(\Delta t_\mathrm{CFL}\) until the full \(\Delta t_\mathrm{sub}\) interval is covered. This mechanism ensures that the hyperbolic component of the solver remains stable even under dynamically varying conditions.
\end{enumerate}

Overall, this multi-level timestep control strategy ensures a consistent and 
stable temporal integration, balancing accuracy, efficiency and the specific 
requirements of each physical process represented in the operator-splitting scheme.

\section{Addressing the reliability of \texttt{SAETASS} for physical modeling}
\label{sec:tests}

To assess the accuracy, robustness and consistency of the numerical scheme, we perform a systematic validation against a set of test problems with known theoretical behaviour. The validation strategy is designed to isolate and verify each physical operator implemented. The source operator is not validated independently, but used to obtain steady-state (SS) solutions when possible. For each case, we consider configurations that admit either closed-form solutions or well-defined asymptotic and conservation properties, allowing for a direct and unambiguous comparison with the numerical results.

The tests are constructed to probe both the transitory and steady-state regimes, as well as to explore the response of the scheme to different configurations, localized sources and time and spatially varying coefficients. Particular emphasis is placed on convergence with resolution or with simulation time, which provides a quantitative measure of the formal order of accuracy of the method and its stability properties. We also aim to study the convergence towards steady-state solutions.

Whenever applicable, error norms and residuals are evaluated to characterize the numerical behaviour in a controlled way. In particular, the relative error that we will refer to in the next subsections is defined as the following $L_2$-norm:
\begin{equation}
\label{eq:L2_norm}
\mathcal{E}_{L_2} = \sqrt{ \frac{\sum_i \left| f_{i}^{\text{num}}(T) - f_{i}^{\text{theo}}(T) \right|^2}{\sum_i \left| f_{i}^{\text{theo}}(T) \right|^2} },
\end{equation}
where the sum runs over all cell grids at a fixed time $T$ relevant to each case, and where $f^{\text{num}}$ and $f^{\text{theo}}$ are the numerical and theoretical solution, respectively. We expect this error to decrease systematically as resolution increases, so as to prove proper convergence.

Hence, this section presents the numerical outcomes of these tests, while the corresponding analytical solutions and theoretical arguments that underlay the validation are collected in Appendix~\ref{app:analytical} for completeness. Some particular solutions are obtained through the method of manufactured solutions (MMS) as it is shown in Appendix~\ref{app:mms_validation}. Together, these results demonstrate that the proposed implementation correctly captures the expected physical behaviour of the transport equation in spherical symmetry across a broad range of regimes relevant for practical applications. The figures and tests presented in the following sections can be reproduced using the scripts within \texttt{/validation} in the official \texttt{SAETASS} repository\footnote{Accessible in \href{https://github.com/jmgarciamorillo/SAETASS}{https://github.com/jmgarciamorillo/SAETASS}.}.

\subsection{Advection validation}
\label{sec:adv_validation}

We first consider a pure advection test with a constant radial velocity in spherical symmetry, where only the advective operator is active. The initial condition consists of a Gaussian profile in the radial coordinate, corresponding to a thin spherical shell. As the profile is advected outwards, its amplitude is expected to decrease due to purely geometrical effects associated with spherical divergence, while preserving its overall shape in the absence of diffusion or sources. This problem admits an analytical solution, which is derived and discussed in Appendix~\ref{app:advection_constant}.

\begin{figure}[ht]
    \centering
    \begin{subfigure}[t]{0.49\linewidth}
        \centering
        \adjustbox{valign=c,max width=\linewidth}{\includegraphics{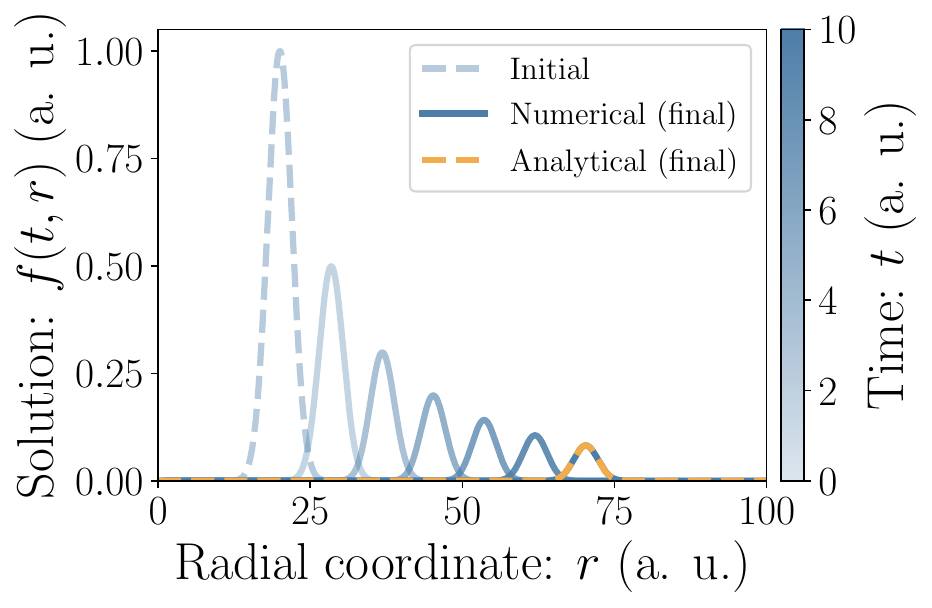}%
        }
        \caption{Comparison between analytical and numerical solution of the purely advective test for the case of $n_r=2^{14}$ radial cells.}
        \label{fig:advection_gaussian_shell_last}
    \end{subfigure}
    \hfill
    \begin{subfigure}[t]{0.49\linewidth}
        \centering
        \adjustbox{valign=c,max width=\linewidth}{\includegraphics{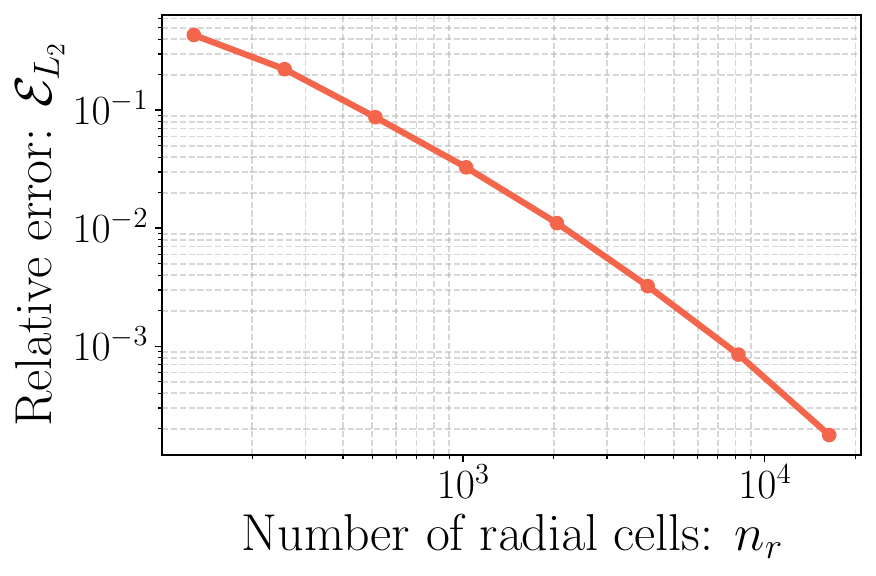}%
        }
        \caption{Convergence study of the relative error for the purely advective test.}
        \label{fig:advection_gaussian_shell_convergence}
    \end{subfigure}

    \caption{Results for the purely advective test. Figure~\ref{fig:advection_gaussian_shell_last} shows qualitative comparison between theoretical and numerical results, whereas Figure~\ref{fig:advection_gaussian_shell_convergence} shows a quantitative convergence analysis.}
    \label{fig:advection_gaussian_shell}
\end{figure}

Figure~\ref{fig:advection_gaussian_shell_last} shows the initial condition together with a comparison between the numerical solution and the analytical result at a representative simulation time for one of the adopted spatial resolutions. In addition, Figure~\ref{fig:advection_gaussian_shell_convergence} displays the convergence behavior of the solver, quantified by measuring the error as a function of resolution. The resulting error as defined in \eqref{eq:L2_norm} exhibits a clear power-law decrease with grid refinement, consistent with the expected order of accuracy of the scheme and confirming the correct implementation of the advection operator in spherical symmetry.

As a second advection test, we consider the radial transport of a localized continuous injection source concentrated in a thin spherical shell around the origin, spanning the radial interval $r \in [0.9,1.1]$. Outside this region, the medium is initially empty and evolves solely under the action of the advection operator in spherical symmetry. This configuration is designed to mimic the behaviour of a point-like source once the advected signal propagates to radii much larger than the source extent.

\begin{figure}[ht]
    \centering
    \begin{subfigure}[t]{0.49\linewidth}
        \centering
        \adjustbox{valign=c,max width=\linewidth}{\includegraphics{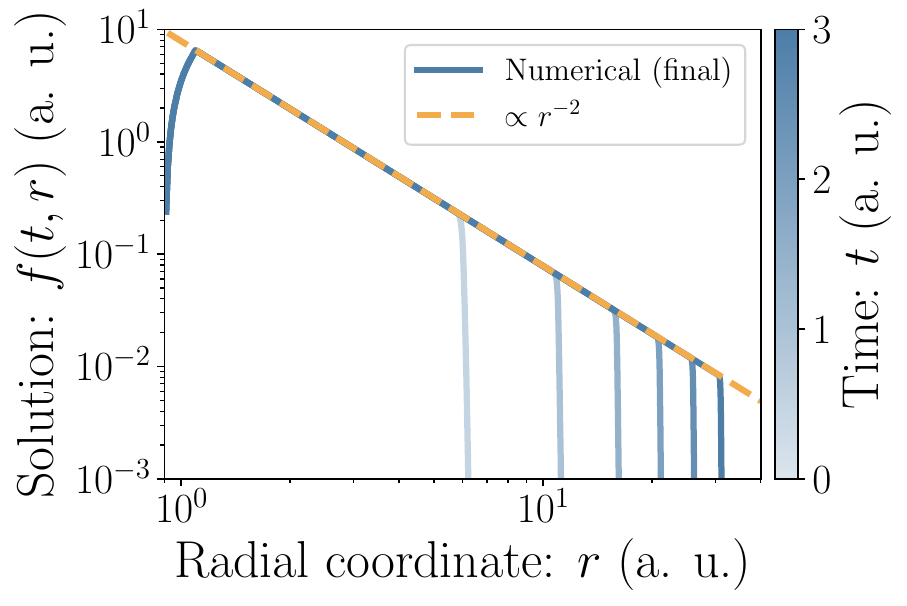}%
        }
        \caption{Time evolution of the pointlike source advection test for $n_r=2^{12}$ radial cells and comparison with the expected $\propto r^{-2}$ trend.}
        \label{fig:advection_pointlike_source_last}
    \end{subfigure}
    \hfill
    \begin{subfigure}[t]{0.49\linewidth}
        \centering
        \adjustbox{valign=c,,max width=\linewidth}{\includegraphics{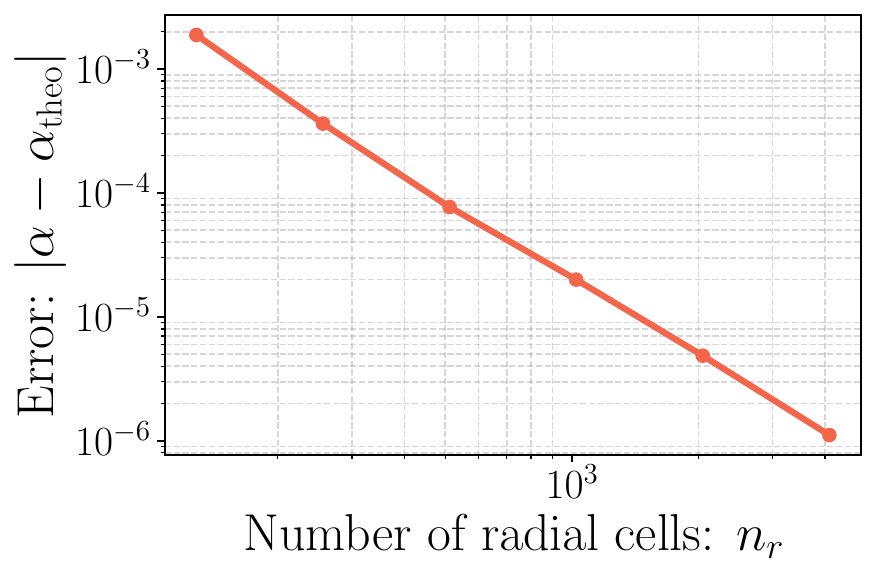}%
        }
        \caption{Convergence of the numerical slope towards the expected $\alpha_\mathrm{theo}=-2$ as the number of radial cells, $n_r$, increases.}
        \label{fig:advection_pointlike_source_convergence}
    \end{subfigure}

    \caption{Results for the advective test with pointlike source. Figure~\ref{fig:advection_pointlike_source_last} shows qualitative comparison between expected and numerical results, whereas Figure~\ref{fig:advection_pointlike_source_convergence} shows a quantitative analysis of the convergence of the observed slope towards the expected one.}
    \label{fig:advection_pointlike_source}
\end{figure}

Figure~\ref{fig:advection_pointlike_source_last} shows the radial profile of the solution at several simulation times, represented in log-log scale. As the advected signal propagates outwards, the numerical solution develops a clear power-law behavior, with a slope consistent with $f(r) \propto r^{-2}$, as expected from geometrical dilution in spherical symmetry for a point-like source under pure advection. For reference, the theoretical $r^{-2}$ trend is overplotted, demonstrating excellent agreement with the numerical results across the sampled radial range and times.

Figure~\ref{fig:advection_pointlike_source_convergence} quantifies this behaviour by measuring the absolute difference between the measured $\alpha$ and theoretical $\alpha_\mathrm{theo} = -2$ slopes of the solution as a function of spatial resolution. This absolute difference vanishes following a power-low trend in $n_r$, as it is expected from the numerical scheme implemented. This confirms that the scheme accurately reproduces the asymptotic behaviour of a point-like source under radial advection in spherical symmetry.

Finally, we consider an advection test designed to illustrate the interplay between geometrical dilution and a spatially varying advection velocity. In this setup, the radial velocity is prescribed to scale as $u_\mathrm{w}(r) \propto r^{-2}$. Under this condition, the decrease in advection speed with radius is expected to exactly compensate the geometrical factor arising from spherical divergence, resulting in a flat stationary radial profile for the purely advected quantity. The detailed derivation of this setup is discussed in Appendix~\ref{app:advection_variable_regularized}.

\begin{figure}[ht]
    \centering
    \begin{subfigure}[t]{0.49\linewidth}
        \centering
        \adjustbox{valign=c,max width=\linewidth}{\includegraphics{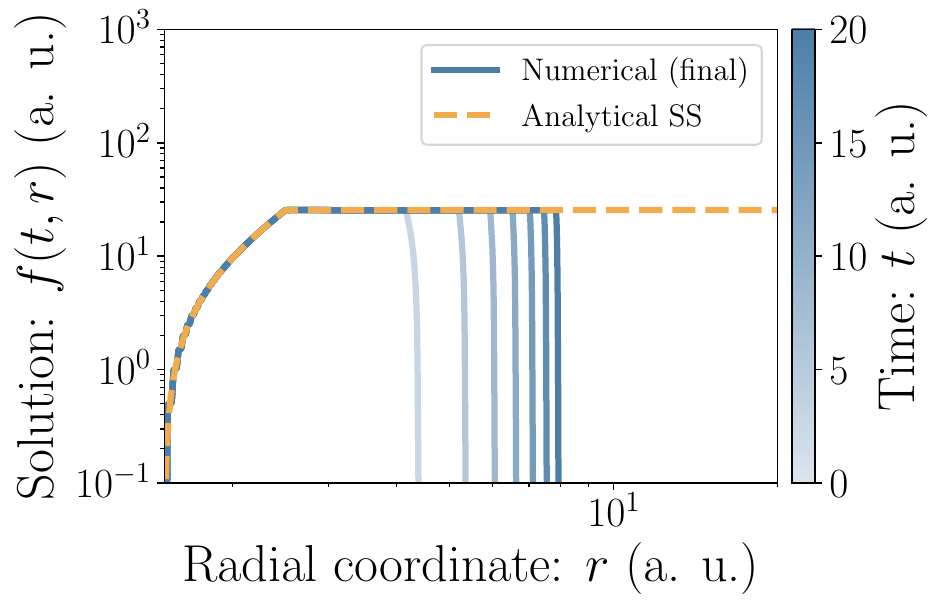}%
        }
        \caption{Time evolution of the distribution towards the analytical steady state for the advection problem with variable velocity for $n_r = 2^{14}$.}
        \label{fig:advection_piecewise_last}
    \end{subfigure}
    \hfill
    \begin{subfigure}[t]{0.49\linewidth}
        \centering
        \adjustbox{valign=c,max width=\linewidth}{\includegraphics{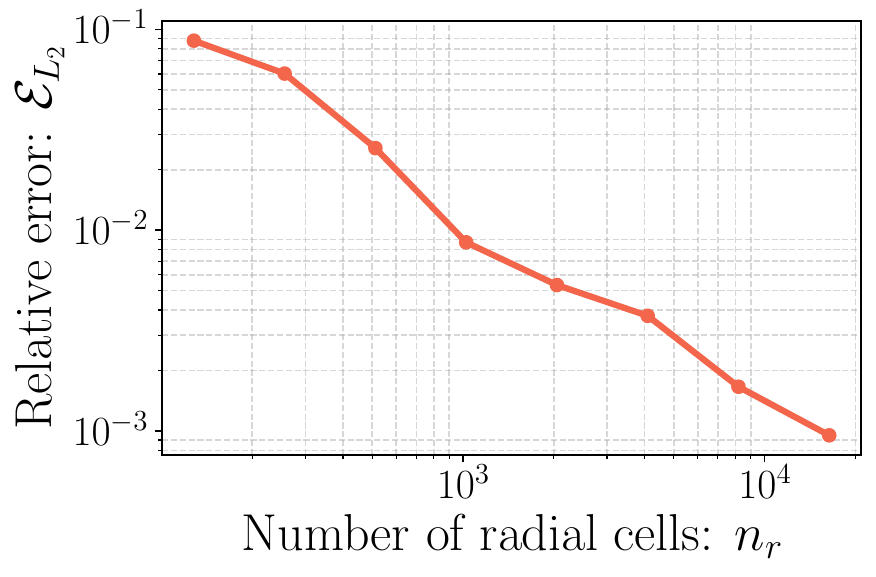}%
        }
        \caption{Convergence of the relative $L_2$ error as a function of the number of spatial grid points, $n_r$.}
        \label{fig:advection_piecewise_convergence}
    \end{subfigure}

    \caption{Time evolution of a locally injected distribution under a $\propto r^{-2}$ advection wind profile. Geometric factors and wind spatial dependence roughly balance out to give a flat distribution.}
    \label{fig:advection_variable_velocity}
\end{figure}

We perform a numerical simulation of this configuration using a localized source and evolve it under the action of the advection operator for different resolutions and for an arbitrary time of simulation chosen in order to properly observe the system evolution. A resulting solution for a specific resolution is shown in Figure~\ref{fig:advection_piecewise_last} in a log-log representation of the radial profile at different simulation times. As expected, the numerical results evolves towards the steady state as time advances.

As shown in Figure~\ref{fig:advection_piecewise_convergence}, the dependence of $\mathcal{E}_{L_2}$ with $n_r$ exhibits clear convergence behaviour, which proves the correct implementation of variable terms in the hyperbolic operator. Note that, in this case, $\mathcal{E}_{L_2}$ is computed including only the region where the solution is non zero, this is, where the particles have had time to arrive.

\subsection{Diffusion validation}
\label{sect:diff_validation}

As a first test for the diffusion operator, we consider a simple constant-diffusion scenario in spherical symmetry. The initial radial profile is chosen as $f_0(r) = \frac{\pi\sin(r)}{2r}$, which admits an exact analytical evolution under the action of a constant diffusion coefficient $D$. The derivation of such solution is presented in Appendix~\ref{app:diffusion_constant}. This setup allows for a straightforward assessment of both the temporal evolution and the spatial accuracy of the numerical scheme.

\begin{figure}[ht]
    \centering
    \begin{subfigure}[t]{0.49\linewidth}
        \centering
        \adjustbox{valign=c,max width=\linewidth}{\includegraphics{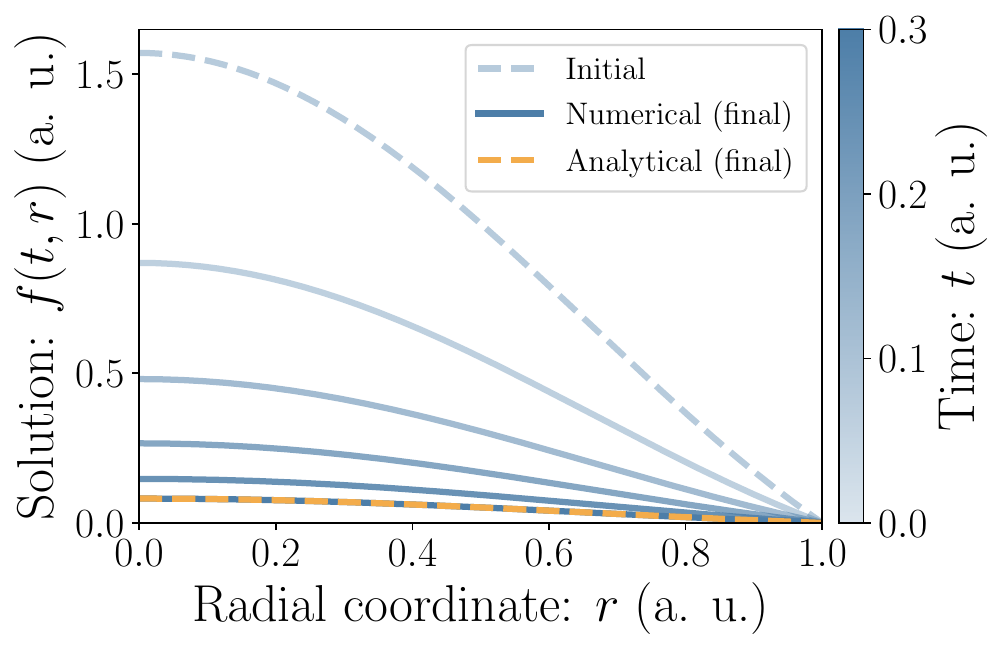}%
        }
        \caption{Comparison between analytical and numerical solution of the purely diffusive test for the case of $n_r=2^{14}$ radial cells.}
        \label{fig:diffusion_analytic_last}
    \end{subfigure}
    \hfill
    \begin{subfigure}[t]{0.49\linewidth}
        \centering
        \adjustbox{valign=c,max width=\linewidth}{\includegraphics{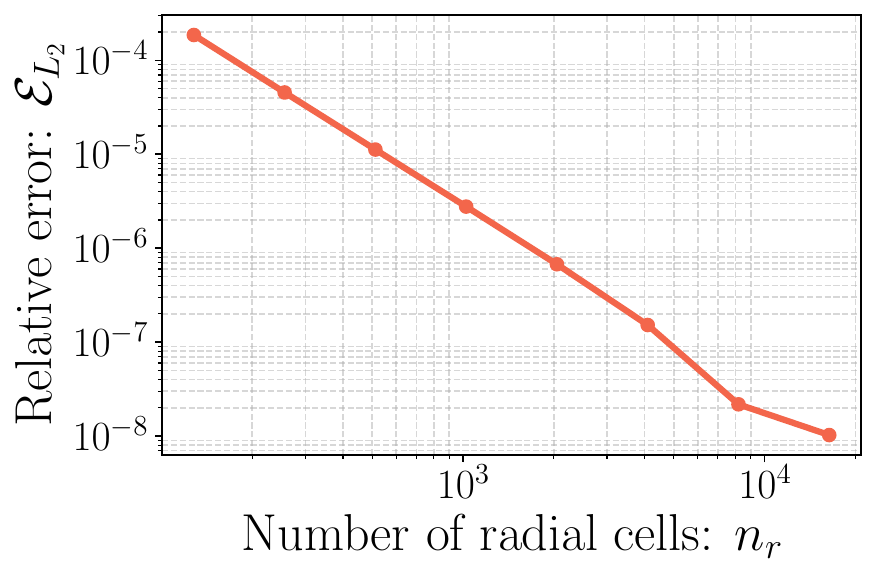}%
        }
        \caption{Convergence study of the relative error for the purely advective test.}
        \label{fig:diffusion_analytic_convergence}
    \end{subfigure}

    \caption{Results for the purely diffusive test. Figure~\ref{fig:diffusion_analytic_last} shows qualitative comparison between theoretical and numerical results, whereas Figure~\ref{fig:diffusion_analytic_convergence} shows a quantitative convergence analysis.}
    \label{fig:diffusion_analytic}
\end{figure}

Figure~\ref{fig:diffusion_analytic_last} shows a comparison between the numerical solution and the analytical result at a representative simulation time. The evolution of the profile is captured accurately, demonstrating that the scheme reproduces the expected smoothing and spreading characteristic of diffusion in spherical symmetry. Figure~\ref{fig:diffusion_analytic_convergence} quantifies the convergence of the solver with increasing spatial resolution. A clear decrease of the error with grid refinement is observed, confirming that the implementation achieves the expected order of accuracy and correctly handles the Laplacian operator in spherical coordinates.

Finally, to evaluate the capability of the scheme to handle non-uniform diffusion coefficients, we consider a steady-state diffusion scenario. Specifically, we prescribe the profiles $D(r) = D_0(\epsilon + r)^2$ and $Q(r) = Q_0 r$, where $\epsilon$ is a small smoothing parameter introduced to avoid singularities at the origin. This configuration admits an exact analytical solution in the steady state, the step-by-step derivation of which is provided in Appendix~\ref{app:diffusion_variable_steady}.

\begin{figure}[ht]
    \centering
    \begin{subfigure}[t]{0.49\linewidth}
        \centering
        \adjustbox{valign=c,max width=\linewidth}{\includegraphics{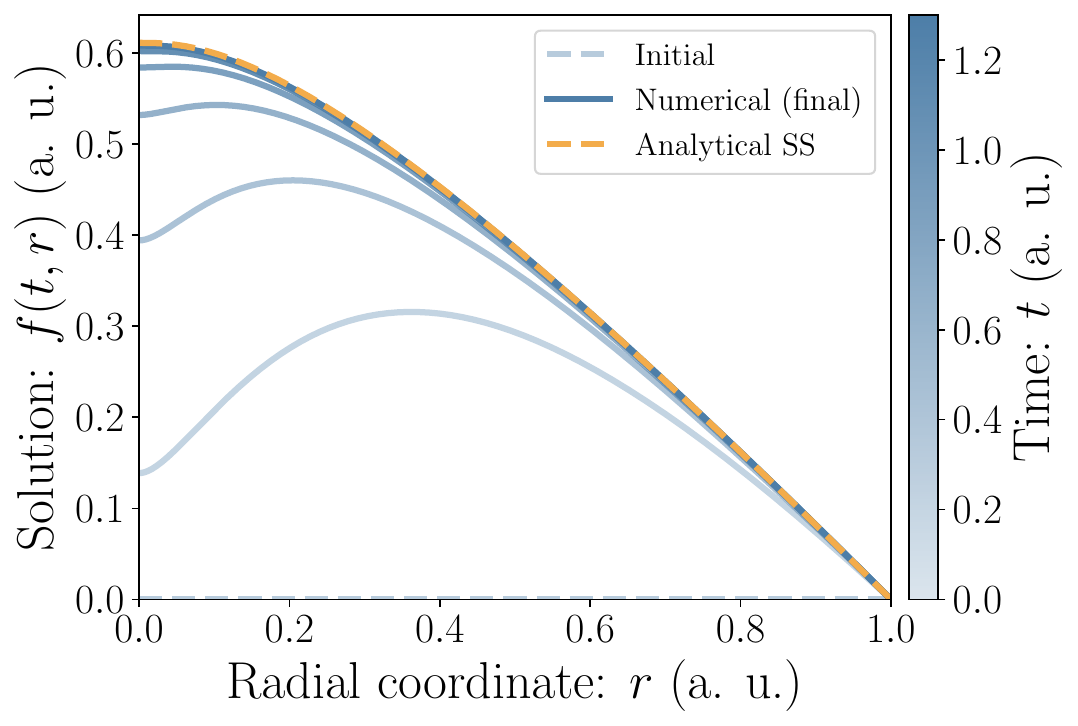}}
        \caption{Time evolution of the distribution towards the analytical steady state for the spatially dependent diffusive problem.}
        \label{fig:diffusion_source_steady_last}
    \end{subfigure}
    \hfill
    \begin{subfigure}[t]{0.49\linewidth}
        \centering
        \adjustbox{valign=c,max width=\linewidth}{\includegraphics{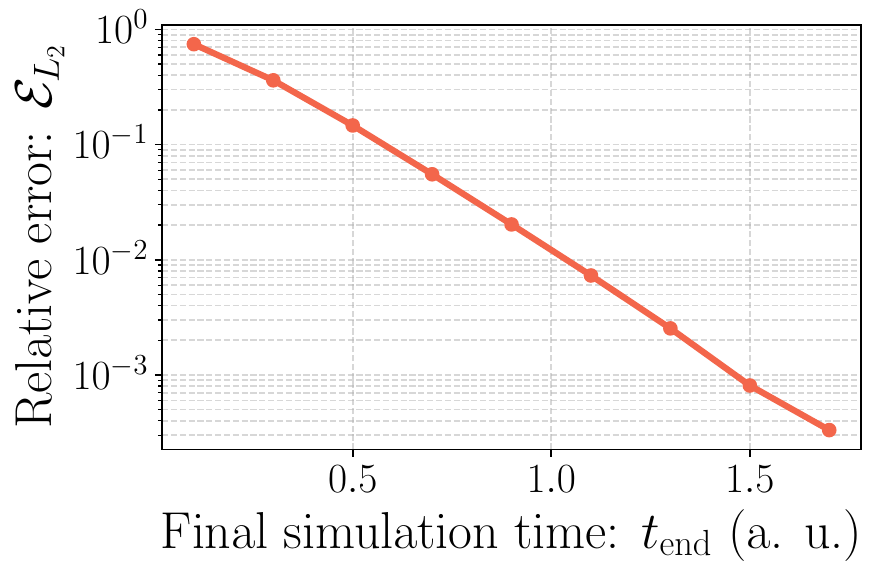}}
        \caption{Convergence of the relative $L_2$ error as a function of the total simulation time $t_{\text{end}}$.}
        \label{fig:diffusion_source_steady_convergence}
    \end{subfigure}

    \caption{Results for the diffusion test with variable coefficients. Figure~\ref{fig:diffusion_source_steady_last} shows the relaxation process towards equilibrium, while Figure~\ref{fig:diffusion_source_steady_convergence} quantifies the accuracy of the reached steady state.}
    \label{fig:diffusion_source_steady}
\end{figure}

The simulations are initialized with a zero distribution and evolved up to a certain $t_\mathrm{end}$. We increase $t_\mathrm{end}$ for each simulation so as to ensure that the system reaches the stationary state. Figure~\ref{fig:diffusion_source_steady_last} illustrates the radial profile $f(t,r)$ at several simulation times. It can be observed how the numerical solution relaxes from the initial condition toward the analytical steady-state profile, correctly reproducing the spatial gradient dictated by the variable diffusivity.

The accuracy of the method in this regime is quantified by measuring the relative $L_2$ error with respect to the final analytical solution. As shown in Figure~\ref{fig:diffusion_source_steady_convergence}, the error exhibits an exponential decay with simulation time, confirming that the numerical scheme is stable under variable coefficients and converges consistently toward the expected theoretical limit.

\subsection{Loss validation}
\label{sec:loss_validation}

As discussed in Section \ref{sec:generalized_variable}, the advection and loss operators are implemented in a combined form, so, in formal terms, no dedicated analysis for the loss component would be needed. However, we present here an explicit validation of the loss term for completeness. The test consists of a non-localized source term combined with a loss term chosen such that the system admits a known analytical steady-state solution, derived in Appendix~\ref{app:loss_steady}. 

\begin{figure}[ht]
    \centering
    \begin{subfigure}[t]{0.49\linewidth}
        \centering
        \adjustbox{valign=c,max width=\linewidth}{\includegraphics{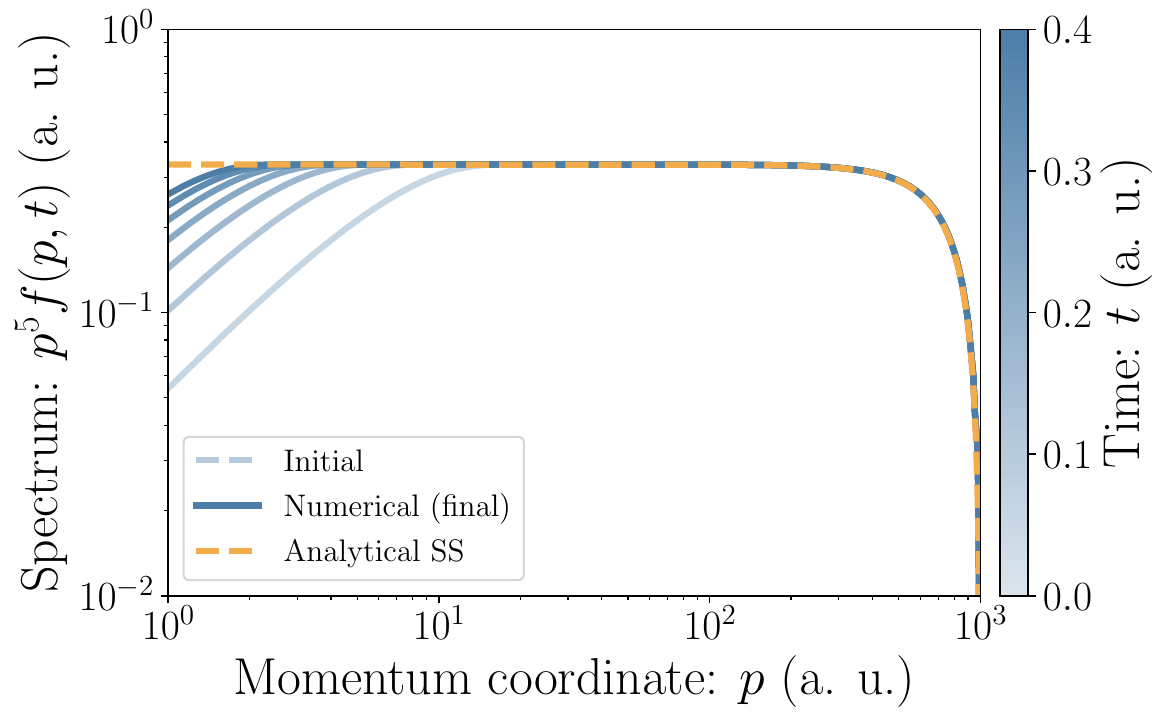}%
        }
        \caption{Time evolution of the distribution towards the analytical steady state for the loss problem.}
        \label{fig:loss_source_steady_last}
    \end{subfigure}
    \hfill
    \begin{subfigure}[t]{0.49\linewidth}
        \centering
        \adjustbox{valign=c,max width=\linewidth}{\includegraphics{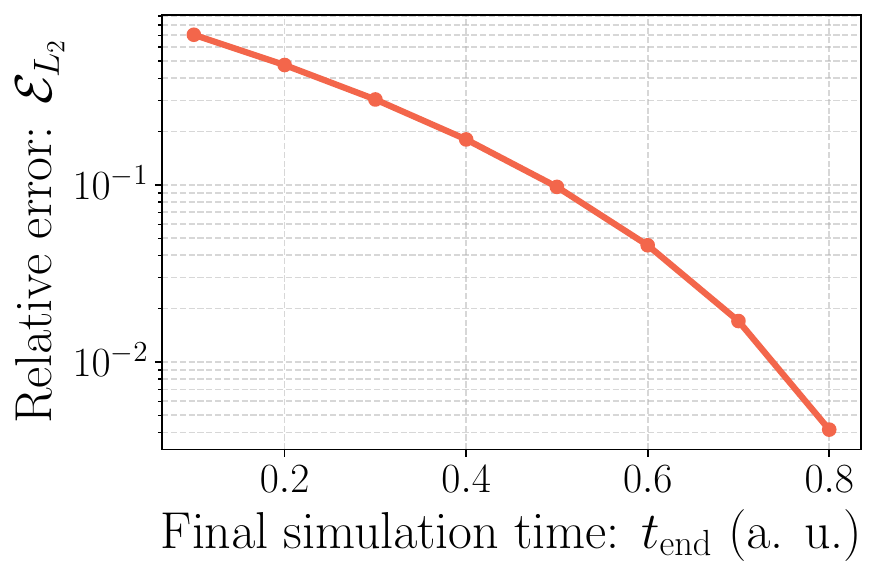}%
        }
        \caption{Convergence of the relative $L_2$ error as a function of the total simulation time $t_{\text{end}}$.}
        \label{fig:loss_source_steady_convergence}
    \end{subfigure}

    \caption{Results for the steady state loss test. Figure~\ref{fig:loss_source_steady_last} shows the relaxation process towards equilibrium, while Figure~\ref{fig:loss_source_steady_convergence} quantifies the accuracy of the reached steady state.}
    \label{fig:loss_source_steady}
\end{figure}

Analogously to what was done in Section~\ref{sect:diff_validation}, the simulations are initialized with a zero distribution and evolved up to a certain $t_\mathrm{end}$. We increase $t_\mathrm{end}$ for each simulation so as to ensure that the system reaches a stationary state. Results are illustrated in Figure~\ref{fig:loss_source_steady}. This demonstrates that the numerical scheme correctly reproduces the balance between source injection and losses in spherical symmetry.

Figure~\ref{fig:loss_source_steady_last} displays the time evolution of a specific run, whereas Figure~\ref{fig:loss_source_steady_convergence} shows the convergence of the solution as a function of the total simulation time $t_\mathrm{end}$. As $t_\mathrm{end}$ increases, the numerical solution asymptotically approaches the steady-state profile, confirming that the scheme correctly captures the temporal evolution towards equilibrium and reproduces the expected analytic steady-state limit.

\subsection{Time dependence validation}
\label{sec:time_validation}

While the previous tests primarily verify the spatial discretization and the steady-state limits, the temporal accuracy of the scheme requires a dedicated assessment, especially when dealing with non-stationary physical conditions. To this end, we employ the method of manufactured solutions (MMS) described in Appendix~\ref{app:mms_validation} for both advection and diffusion operators with explicit time-dependent coefficients. As previously mentioned, loss solver inherits from exactly the same numerical implementation that advection solver, so no further specific loss test will be performed in this section.

These validations are also performed in two stages as in previous sections. First, a qualitative comparison is made by examining the radial profiles at several snapshots. This test visually confirms that the solver correctly captures the behaviour of the system. Second, we perform a quantitative convergence study to verify the formal order of accuracy of the temporal integrator, using the same definition given in \eqref{eq:L2_norm}.

For the time-dependent advection problem, we manufacture a system where the target distribution, $f_\mathrm{MMS}$, oscillates in time while decaying exponentially in the radial direction. This distribution is subjected to a velocity field that expands linearly with radius but slows down over time. The specific required source term, $Q_\mathrm{MMS}$, to sustain this exact evolution is obtained in Appendix~\ref{app:advection_mms}.

Figure~\ref{fig:time_dependent_advection} corresponds to this advective setup. Figure \ref{fig:time_dependent_advection_last} demonstrates the good qualitative agreement between the numerical and analytical solutions after some relevant time of simulation. Figure \ref{fig:time_dependent_advection_convergence} displays a systematic convergence study in space and time. 

\begin{figure}[ht]
    \centering
    \begin{subfigure}[t]{0.49\linewidth}
        \centering
        \adjustbox{valign=c,max width=\linewidth}{\includegraphics{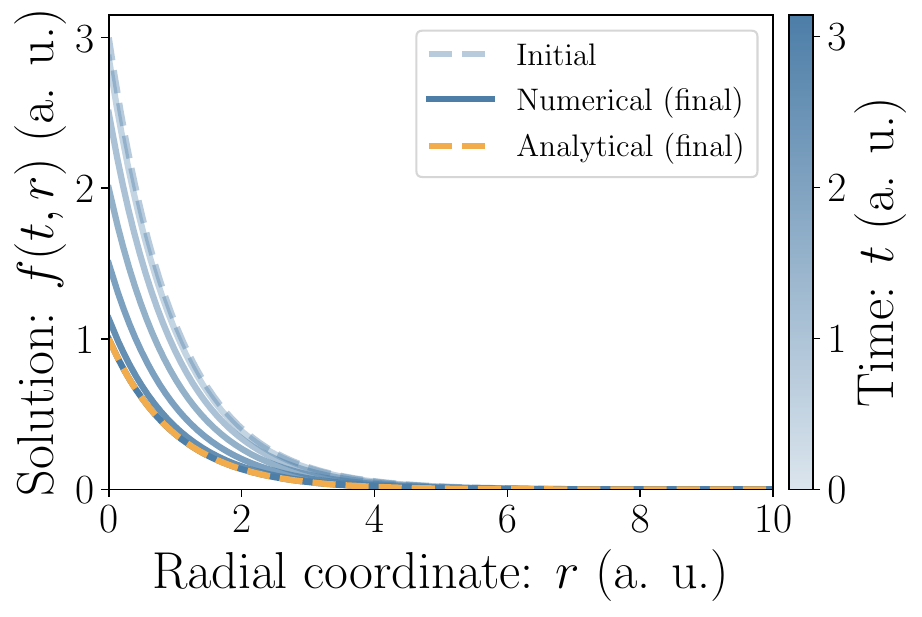}%
        }
        \caption{Time evolution of the distribution and comparison with theoretical solution at a relevant time for the time dependent advective problem and for $n_r=2^{14}$ radial cells.}
        \label{fig:time_dependent_advection_last}
    \end{subfigure}
    \hfill
    \begin{subfigure}[t]{0.49\linewidth}
        \centering
        \adjustbox{valign=c,max width=\linewidth}{\includegraphics{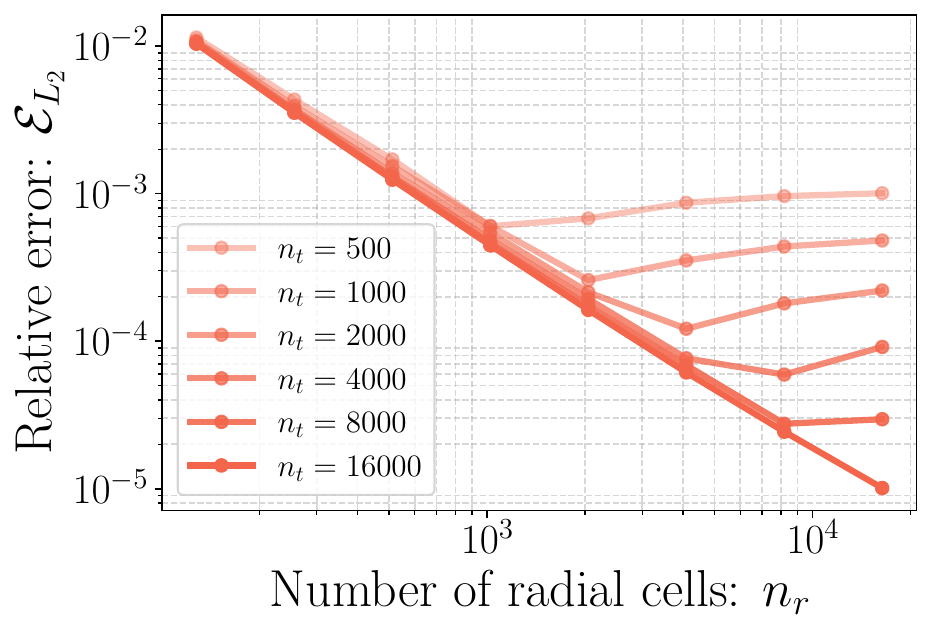}%
        }
        \caption{Convergence of the relative $L_2$ error as a function of the number of radial cells, $n_r$. Different convergence curves are displayed for different number of temporal substeps, $n_t$.}
        \label{fig:time_dependent_advection_convergence}
    \end{subfigure}

    \caption{Results for the time dependent advective problem. Figure~\ref{fig:time_dependent_advection_last} shows the comparison between numerical and analytical solutions. Figure~\ref{fig:time_dependent_advection_convergence} quantifies the type of convergence acquired in time and space.}
    \label{fig:time_dependent_advection}
\end{figure}

It is remarkable how Figure~\ref{fig:time_dependent_advection_convergence} showcases the interplay between temporal and spatial resolution. Indeed, temporal resolution establishes the minimum relative error achievable by the numerical scheme, under which further spatial refinement cannot lead.

\begin{figure}[ht]
    \centering
    \begin{subfigure}[t]{0.49\linewidth}
        \centering
        \adjustbox{valign=c,max width=\linewidth}{\includegraphics{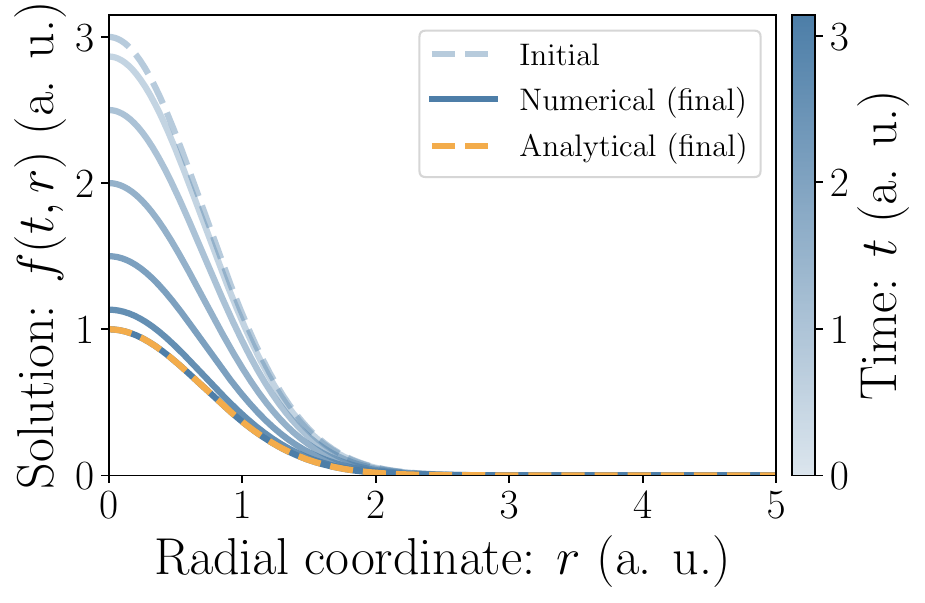}%
        }
        \caption{Time evolution of the distribution and comparison with theoretical solution at a relevant time for the time dependent diffusion problem and for $n_r=2^{14}$ radial cells.}
        \label{fig:time_dependent_diffusion_last}
    \end{subfigure}
    \hfill
    \begin{subfigure}[t]{0.49\linewidth}
        \centering
        \adjustbox{valign=c,max width=\linewidth}{\includegraphics{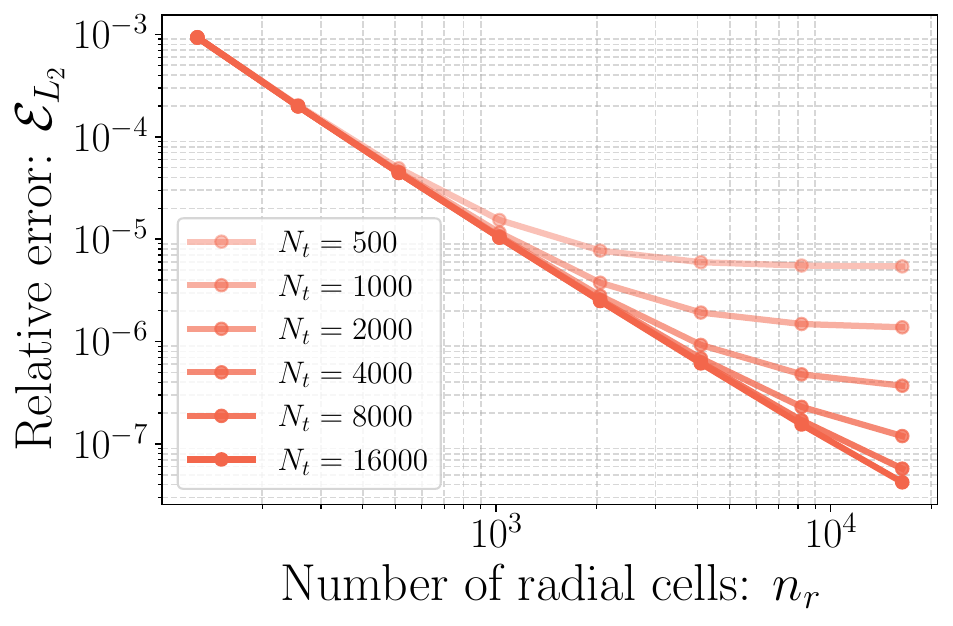}%
        }
        \caption{Convergence of the relative $L_2$ error as a function of the number of radial cells, $n_r$. Different convergence curves are displayed for different number of temporal substeps, $n_t$.}
        \label{fig:time_dependent_diffusion_convergence}
    \end{subfigure}

    \caption{Results for the time dependent diffusive problem. Figure~\ref{fig:time_dependent_diffusion_last} shows the comparison between numerical and analytical solutions. Figure~\ref{fig:time_dependent_diffusion_convergence} quantifies the type of convergence acquired in time and space.}
    \label{fig:time_dependent_diffusion}
\end{figure}

Analogous analysis is displayed in Figure~\ref{fig:time_dependent_diffusion}, for the time dependent diffusive problem. In this case, we construct a target distribution featuring an oscillatory time dependence coupled with a Gaussian spatial decay. This distribution evolves under a spatially uniform diffusion coefficient, $D(t)$, that increases linearly with time. The exact analytical forms for this setup are provided in Appendix~\ref{app:diffusion_mms}.

Similar results emphasize the correctness of the implementation of the different numerical routines for the different physical solvers.

\section{Application to a realistic astrophysical system}
\label{sec:applications}

In order to demonstrate the versatility and accuracy of \texttt{SAETASS} package, we apply it to a physical scenario discussed in \citep{Morlino2021} and valid for compact star clusters\footnote{We here define compact star clusters as those where stars are close enough for their stellar winds to merge and form a collective outflow.}. The authors modelled the transport of cosmic ray protons within a stellar cluster bubble using the classic wind-blown bubble framework presented by Weaver in \citep{Weaver1977}. 

While the analytical approach in \citep{Morlino2021} provides valuable insight, it is limited to steady-state solutions and requires the neglect of energy losses (except from the adiabatic ones) in order to achieve a solvable problem from the mathematical point of view. In contrast, \texttt{SAETASS} solves the full time-dependent transport equation in spherical symmetry and also allows for the inclusion of arbitrary diffusion coefficients or loss mechanisms.

\begin{figure}[ht]
    \centering
    \includegraphics[width=\linewidth]{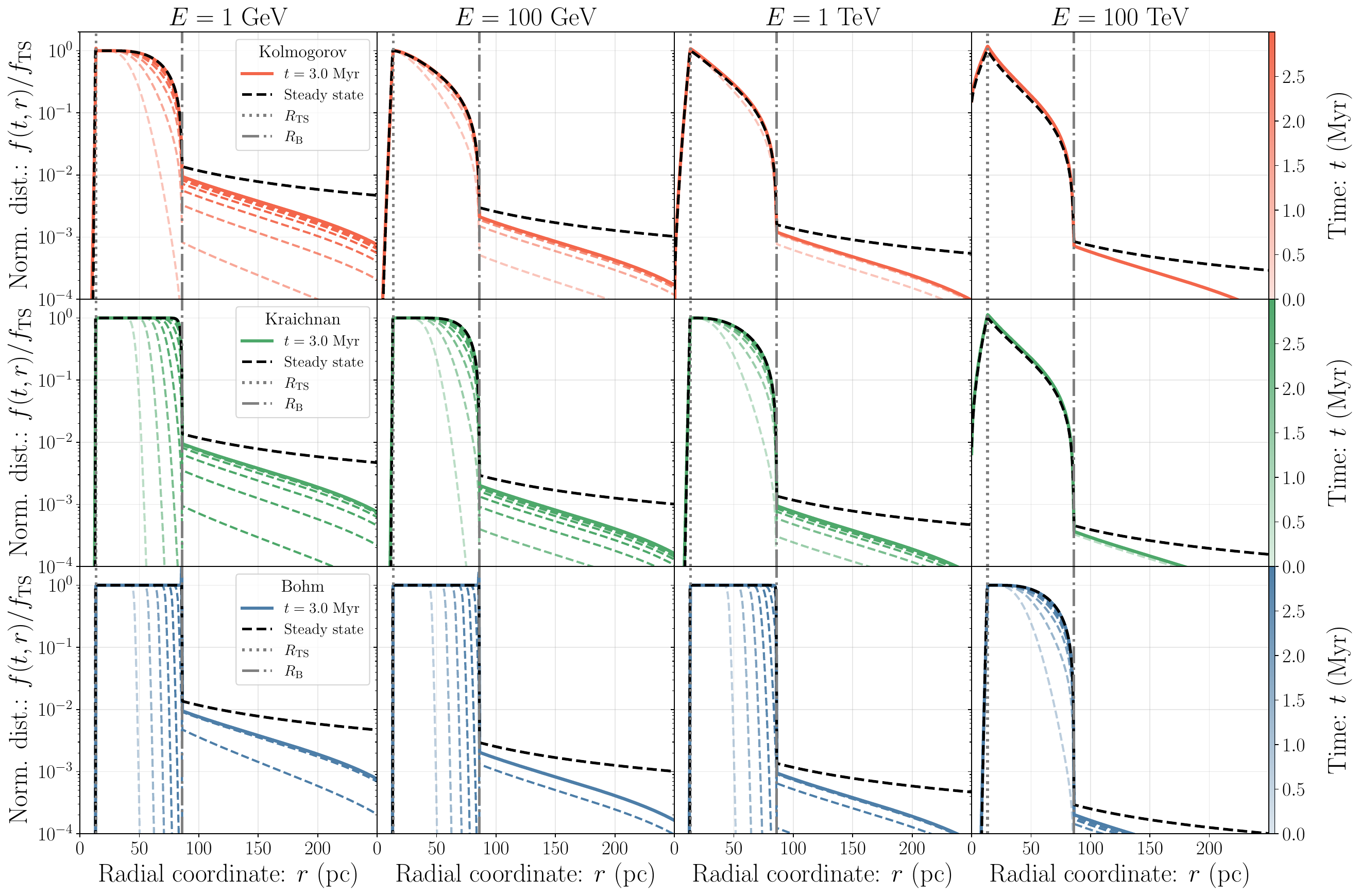}
    \caption{Radial distribution of the normalized cosmic ray proton distribution function $f(t,r)$ at different energies ($E = 1, 100$~GeV, $1, 100$~TeV) for three different diffusion models: Kolmogorov (top), Kraichnan (middle) and Bohm (bottom). The colour gradient indicates the temporal evolution from $t=0$ to $t=3$~Myr. The dashed black line represents the analytical steady-state solution \citep{Menchiari2024}. Vertical lines indicate the position of the termination shock ($R_{\rm TS}$) and the bubble radius ($R_{\rm B}$).}
    \label{fig:menchiari_benchmark}
\end{figure}

Figure~\ref{fig:menchiari_benchmark} shows the results of simulations for four different particle energies ($1$~GeV, $100$~GeV, $1$~TeV and $100$~TeV) and three distinct diffusion regimes: Kolmogorov, Kraichnan and Bohm. With the aim of comparing our results with the one displayed in \citep{Menchiari2024}, the simulations are performed with the parameters of the cluster used in that study: $L_\mathrm{w} = 2\cdot10^{38}\ \text{erg s}^{-1}$ as the wind luminosity, $\dot{M} = 10^{-4}\ M_{\odot}\ \text{s}^{-1}$ as the mass loss rate, $t_\mathrm{age}=3\ \text{Myr}$ as the age and $\rho_0 = 20\ m_p\ \text{cm}^{-3}$ as the surrounding ISM mass density. As in \citep{Menchiari2024}, we also neglect energy loss in our calculations. Further information about the script configuration and simulation setup can be found in the \texttt{/examples} directory of \texttt{SAETASS} official repository.

Regarding the choice of total time of simulation, it is interesting to discuss the different evolution timescales of the system: the advective, $\tau_\mathrm{adv}$, and diffusive, $\tau_\mathrm{diff}$, timescales. As advection parameters are independent on the energy of the particles, depending only on the cluster parameters, $\tau_\mathrm{adv}$ is fixed and results an upper limit for the total timescale of evolution of the system, $\tau_\mathrm{tot}$. Hence, $\tau_\mathrm{tot}$ is only modified for sufficiently energetic particles, where diffusion becomes more efficient and can decrease this total timescale. Consequently, we only need to ensure our simulation time is above $\tau_\mathrm{adv}$, which, for our parameters, results in \citep{Menchiari2024}
\begin{equation*}
    \tau_\mathrm{adv} = \frac{4R_\mathrm{TS}}{3u_\mathrm{w}}\left(\frac{R_\mathrm{B}^3}{R_\mathrm{TS}^3}-1\right)=\frac{4R_\mathrm{TS}}{3}\sqrt{\frac{\dot{M}}{2L_\mathrm{w}}}\left(\frac{R_\mathrm{B}^3}{R_\mathrm{TS}^3}-1\right) \approx 1.81\ \text{Myr}.
\end{equation*}
We take $t_\mathrm{end}=3\ \text{Myr}$ and verify, \emph{a posteriori}, that this simulation time leads, indeed, to the steady state of the system.

In these conditions, Figure~\ref{fig:menchiari_benchmark} displays then the time evolution of the spatial distribution of the normalized cosmic ray proton density $f(t,r)/f_{\rm TS}$ as a function of the radial coordinate $r$. In the steady-state regime, the analytical solution provided by \citep{Morlino2021} is indeed obtained. A key strength of \texttt{SAETASS} is its ability to track the temporal evolution towards such steady state. As shown in the figure, the solutions for $t = 3$~Myr almost perfectly overlap with the dashed lines representing the steady-state limit showed in \citep{Menchiari2024}. This benchmark confirms that our numerical scheme correctly handles the sharp transitions at the termination shock ($R_{\rm TS}$) and the bubble radius ($R_{\rm B}$), where the diffusion properties and flow velocities change abruptly.

Furthermore, temporal resolution of \texttt{SAETASS} reveals critical transport dynamics that are invisible in steady-state models. Specifically, Figure~\ref{fig:menchiari_benchmark} illustrates the evolution of the proton distribution across several orders of magnitude in energy. It is evident that high-energy protons ($100$~TeV) reach the steady-state configuration almost instantaneously across the entire spatial domain. This is a direct consequence of their large diffusion coefficients, allowing them to rapidly cross through the geometry and escape the bubble. 

Conversely, for lower energy protons ($1$~GeV), the transport becomes increasingly dominated by advection from the stellar wind. This shift is even more pronounced as we move from Kolmogorov to Bohm diffusion models, where the reduced diffusion coefficients lead to much longer residence times within the bubble. In these cases, the filling of the region beyond the termination shock is significantly slower and the system requires a much larger timescale to converge toward the steady-state limit. \texttt{SAETASS} accurately captures this transition.

\begin{figure}[ht]
    \centering
    \begin{subfigure}[t]{0.49\linewidth}
        \centering
        \adjustbox{valign=c,max width=\linewidth}{\includegraphics{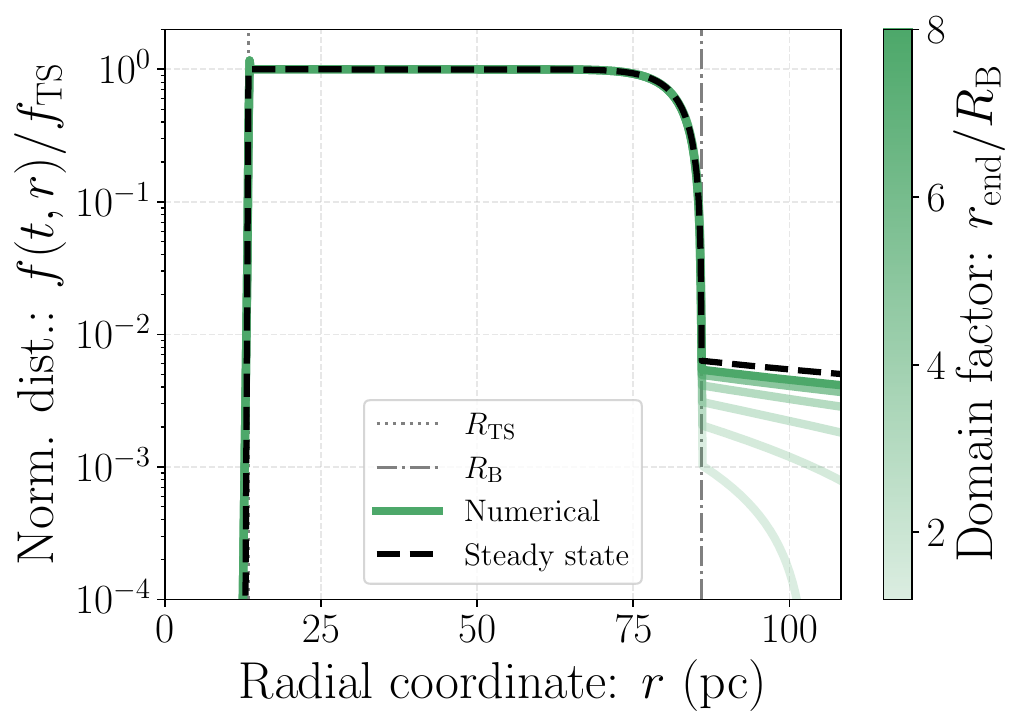}%
        }
        \caption{Results for simulations with different domain factors, $r_\mathrm{end}/R_\mathrm{B}$. It is proven that the behaviour outside $R_\mathrm{B}$ converges to the one in the steady-state solution as $r_\mathrm{end}$ increases.}
        \label{fig:domain_size_solutions}
    \end{subfigure}
    \hfill
    \begin{subfigure}[t]{0.49\linewidth}
        \centering
        \adjustbox{valign=c,max width=\linewidth}{\includegraphics{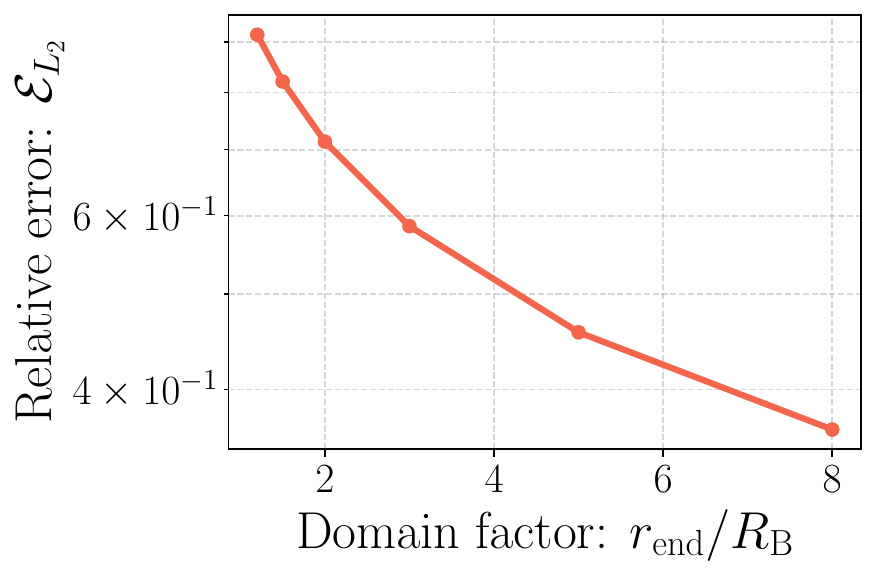}%
        }
        \caption{Convergence of the relative error between numerical and theoretical solutions as defined in equation~\eqref{eq:L2_norm}.}
        \label{fig:domain_size_convergence}
    \end{subfigure}

    \caption{Analysis of the influence in the simulation result of the distance between the boundary and the system. For the simulations, the case of $10$~GeV and Kraichnan diffusion is used. It is shown how the discrepancies between numerical and theoretical solutions inside the ISM in Figure~\ref{fig:menchiari_benchmark} is due to domain size. However, it is proven that this discrepancy does not affect the rest of the solution.}
    \label{fig:domain_size_test}
\end{figure}

The discrepancies observed for the tendency as $r$ increases and leaves the bubble are due to the boundary condition chosen to perform the analysis, this is, $f(t,r_\text{end})=0$; where we took $r_\text{end}=300\ \text{pc}$. Of course, a boundary condition imposing the correct analytical expected value at the boundary could have been imposed, but this would have introduced \textit{a priori} information about the solution. By imposing a zero Dirichlet condition over a far boundary sphere, we emulate the dilution of cosmic ray distribution without affecting the dynamics of the relevant system. Our analysis has proven that the numerical solution converges to that of \citep{Morlino2021, Menchiari2024} as $r_\text{end}\to\infty$, even in the outer region, as shown in Figure~\ref{fig:domain_size_test}. In particular, for the case of $10$~GeV and Kraichnan diffusion, Figure~\ref{fig:domain_size_solutions} qualitatively shows how the ISM tail of the simulation result approaches the expected trend as domain size increases; whereas Figure~\ref{fig:domain_size_convergence} displays a convergence analysis following the style of Section~\ref{sec:tests}.

As we have shown, \texttt{SAETASS} introduces, then, a powerful tool to move beyond analytical simplifications of current state-of-the-art models. While the benchmark in Figure~\ref{fig:menchiari_benchmark} validates the code against the lossless analytical case, extensive analysis to include different energy loss mechanisms can be performed. Additionally, time-dependent profiles as growing bubbles and evolving shocks could also be easily studied with \texttt{SAETASS} current features.

\section{Unlocking time-dependent spherical cosmic ray transport with \texttt{SAETASS}}
\label{sec:conclusions}

In this work, we presented \texttt{SAETASS}, a novel open-source numerical solver designed for the time-dependent astroparticle transport equation in one-dimensional spherical symmetry. By combining a high-resolution, shock-capturing MUSCL-Hancock scheme for hyperbolic advection with an implicit, batched Crank-Nicolson algorithm for parabolic diffusion, the code ensures global particle conservation and numerical stability across steep physical gradients.

The physical accuracy of these numerical schemes was rigorously verified through a suite of tests. Furthermore, the practical application of \texttt{SAETASS} to a wind-blown bubble scenario demonstrated its capability to handle complex, physically motivated setups. The solver seamlessly resolved sharp transitions at shock boundaries and successfully recovered both the pre-equilibrium temporal dynamics and the analytical steady-state profiles of proton transport across different diffusion regimes.

These demonstrated capabilities make \texttt{SAETASS} a highly valuable tool for investigating localized, radially stratified environments. Systems such as expanding supernova remnants, stellar wind-blown bubbles, young pulsar wind nebulae or TeV halos around neutron stars can now be modelled with a fully time-dependent and loss-inclusive numerical treatment. Applications to these specific scenarios will be explored in subsequent works.

While the current version provides a robust and modular foundation, several developments are planned to expand its feature set in upcoming releases. Key future improvements include computational optimizations, such as parallelization, migration to lower-level core routines and adaptive mesh refinements. Additionally, we plan to implement high-order time integrators, flexible coupling interfaces for dynamically evolving background profiles and a dedicated utility subpackage for built-in gamma-ray and neutrino emission calculations.

Ultimately, \texttt{SAETASS} provides the astroparticle physics community with a dedicated, lightweight and rigorous tool, facilitating new studies of particle acceleration and propagation in some of the most extreme environments in the Universe.


\appendix
\newpage

\section{Software design of \texttt{SAETASS} package}
\label{app:software_design}

\begin{wrapfigure}{r}{0.5\textwidth}
    \centering
    \vspace{-1.5cm}
    \begin{tikzpicture}%
  \draw[color=black!60!white]
  \FTdir(\FTroot,root,SAETASS){
  
    \FTdir(root,docs,docs){
      \FTdir(docs,source,source){
      \FTfile(source,[documentation])
      }
      \FTfile(docs,...)
    }
    
    ++(0,-0.5em)
    
    \FTdir(root,src,src) {
      \FTdir(src,saetass,saetass){
        \FTdir(saetass,cli,cli)
        \FTdir(saetass,solvers,solvers){
          \FTfile(solvers,advection\_solver.py)
          \FTfile(solvers,diffusion\_solver.py)
          \FTfile(solvers,...)
        }
        \FTdir(saetass,utils,utils){
          \FTfile(utils,energy\_losses.py)
          \FTfile(utils,bubble\_profiles.py)
          \FTfile(utils,...)
        }
        \FTfile(saetass,grid.py)
        \FTfile(saetass,solver.py)
        \FTfile(saetass,splitting.py)
        \FTfile(saetass,state.py)
      }
    }
    
    ++(0,-0.5em)
    
    \FTdir(root,test,test){
      \FTfile(test,[unit tests])
      \FTfile(test,...)
    }

    ++(0,-0.5em)

    \FTdir(root,tutorials,tutorials){
      \FTfile(tutorials,[tutorial notebooks])
      \FTfile(tutorials,...)
    }

    ++(0,-0.5em)
    
    \FTdir(root,examples,examples){
      \FTfile(examples,[example scripts])
      \FTfile(examples,...)
    }
    
    ++(0,-0.5em)

    \FTdir(root,validation,validation){
      \FTdir(validation,cases,cases){
        \FTfile(cases,[validation scripts])  
      }
      
      \FTfile(validation,...)
    }

    ++(0,-0.5em)
    \FTfile(root,pyproject.toml)
    \FTfile(root,...)
  };
\end{tikzpicture}
    \caption{Overview of the \texttt{SAETASS} repository structure, highlighting the separation between the core engine (\texttt{src/saetass}), physical utilities, validation suites, and documentation.}
    \vspace{-1.5cm}
    \label{fig:repo_structure}
\end{wrapfigure}
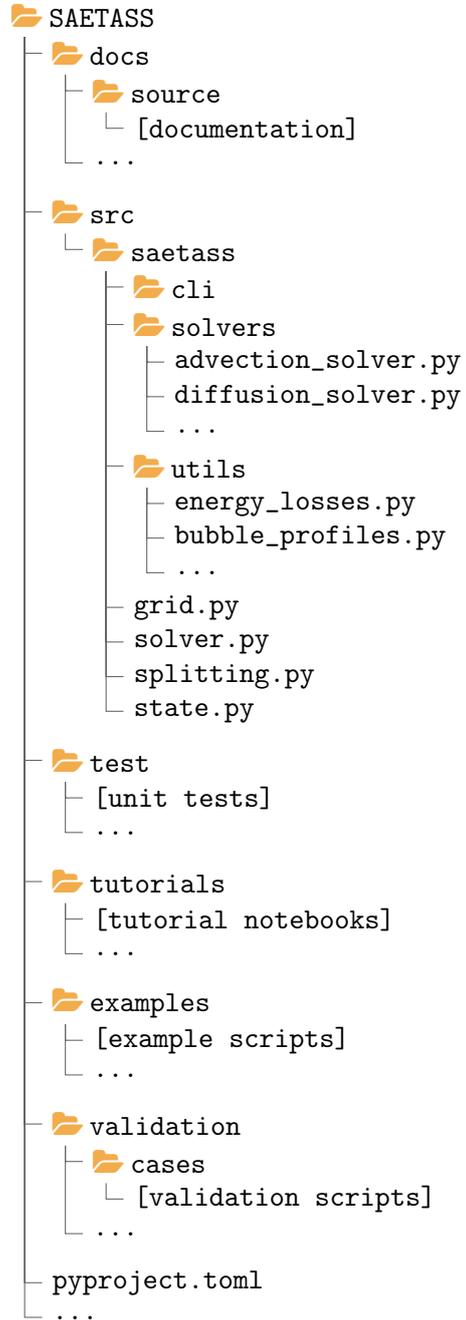

In this appendix we discuss the design of the package structure and the calculation pipeline of \texttt{SAETASS}. This section aims, therefore, to serve as an overview of the specific software implementation performed, in order to summarize the behaviour and usage of \texttt{SAETASS} at its first version. Updated information of these and further features of the package can be found on the \texttt{SAETASS} official repository and documentation.

\subsection{Package structure and architecture}
\label{app:package_structure}

The architecture of \texttt{SAETASS} is explicitly designed to reflect the mathematical operator-splitting approach detailed in Section~\ref{sect:time_discretization_operator_splitting}. By adopting a modular, object-oriented paradigm, the framework ensures that individual physical processes can be modified, removed or expanded without altering the core time-integration engine. This abstraction guarantees that the software remains maintainable and flexible as new numerical schemes or astrophysical environments are incorporated.

As illustrated in Figure~\ref{fig:repo_structure}, the repository is divided into several ecosystems. Beyond the core mathematical engine, the package includes supplementary modules to enhance usability and ease of development. For instance, the \texttt{cli} subpackage is responsible for formatting terminal outputs and managing progress bars handling structured logging. 

Additionally, the \texttt{utils} subpackage provides a suite of ready-to-use physical and astrophysical tools that facilitate the setup of complex simulations. Currently, this includes \texttt{energy\_losses}, an extensive energy loss rates and timescales calculator, which computes hadronic and leptonic interactions such as pion production, Bremsstrahlung, Inverse Compton scattering, etc., as detailed in Appendix~\ref{app:loss_functions_implemented}. The \texttt{utils} subpackage also features the \texttt{bubble\_profiles} module, which offers analytical wind-blown bubble profiles functionalities. This utility suite is designed to grow continuously as new features and astrophysical environments are modelled in the future.

The core of the repository, which drives the numerical integration, resides directly within the \texttt{/src/saetass} directory and its \texttt{/solvers} subdirectory. The logic of this engine revolves around a few key abstract components. First, \texttt{grid.py} and \texttt{state.py} modules manage the physical data structures, handling the spatial and momentum coordinates alongside the instantaneous physical state of the distribution function. The module \texttt{splitting.py} orchestrates the temporal substepping logic. It implements the sequence in which the individual physical operators are applied, as was explained in Section~\ref{sect:time_discretization_operator_splitting} and Section~\ref{sec:substep_calculation}. Besides, \texttt{solver.py} acts as the main orchestrator of the simulation, managing the global advancement of the system. It delegates the specific numerical operations to the individual modules inside \texttt{solvers/}, which implement the specific numerical routines.

For a more concrete understanding of how these core scripts interact during a simulation, see Appendix~\ref{app:typical_workflow}, which illustrates a typical \texttt{SAETASS} workflow, including a detailed sequence diagram of the initialization and time-integration phases.

It is important to note that this section provides only a high-level, general overview of the repository's architecture. The software contains and will develop many more specificities, internal controls and advanced features. For highly detailed and up-to-date information regarding the \texttt{SAETASS} API, class structures and advanced configurations, users are strongly encouraged to consult the official documentation.

\subsection{Simulation workflow in \texttt{SAETASS} package}
\label{app:typical_workflow}

To ensure modularity, maintainability and robust mathematical integration, the software strictly separates the problem configuration from the numerical execution. This separation of concerns is managed by a series of specialized classes, namely \texttt{State}, \texttt{Solver}, \texttt{SplittingScheme}, \texttt{SubSolver} and the \texttt{cli} module; which interact sequentially to advance the system in time.

\begin{figure}[ht]
    \centering
    \includegraphics[width=\textwidth, trim=20 10 20 10, clip]{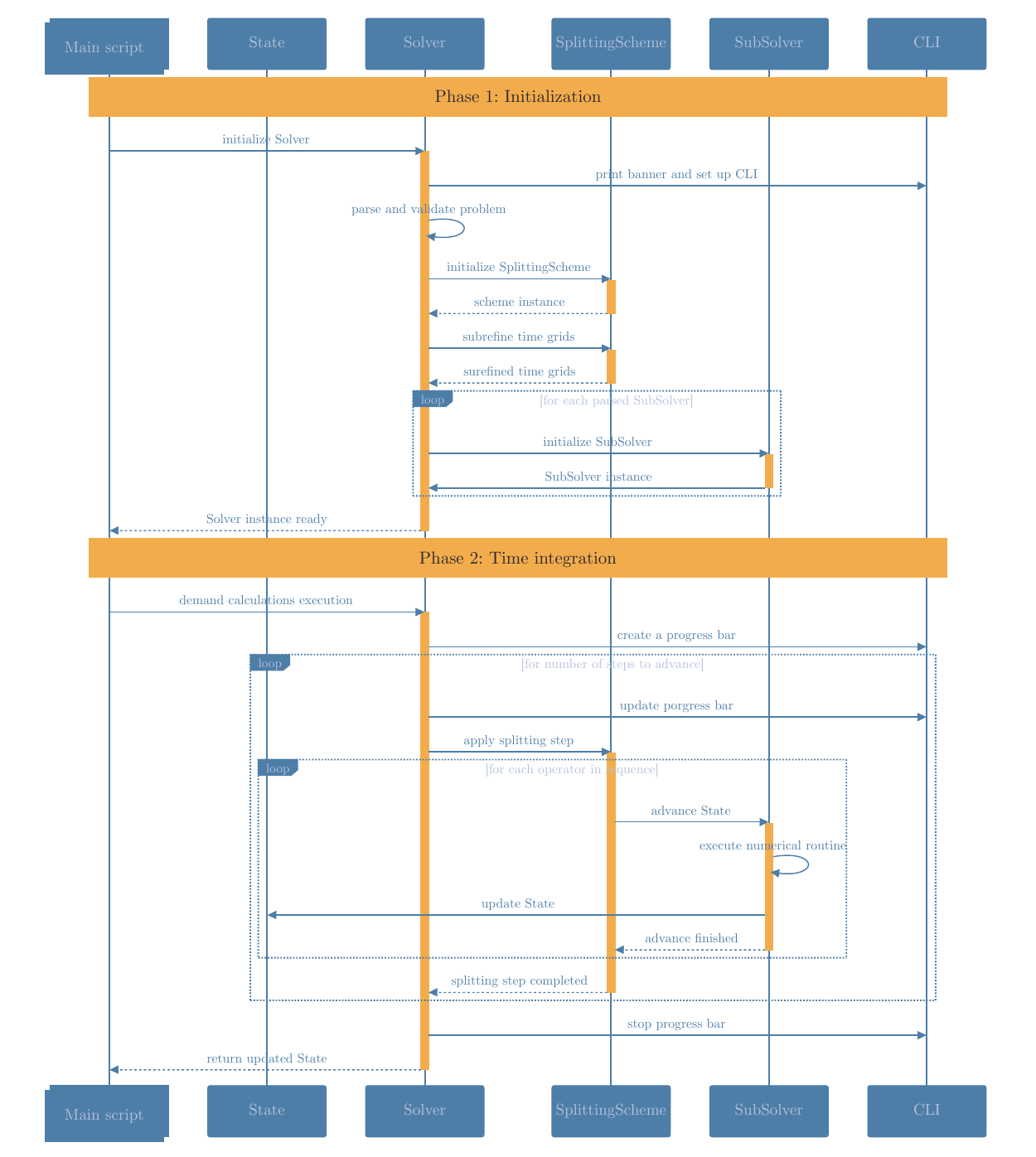}
    \caption{Sequence diagram of a typical workflow of \texttt{SAETASS} calculations. The diagram illustrates the ordered interactions between the main script, the core classes and the command-line interface across the initialization and integration phases.}
    \label{fig:sequence_diagram}
\end{figure}

As illustrated in Figure \ref{fig:sequence_diagram}, the standard execution lifecycle of a \texttt{SAETASS} simulation is divided into two distinct procedural blocks: a phase of initialization and a time integration phase.

First, the initialization phase is responsible for setting up the computational environment, instantiating needed objects and validating the physical parameters before any computationally expensive operations begin. 

The workflow is triggered when a script instantiates the central \texttt{Solver} object. Immediately upon instantiation, the \texttt{Solver} interfaces with the \texttt{cli} module to print the package banner and configure standard output formatting. Following this, the \texttt{Solver} parses and validates the problem definition provided by the user and proceeds to initialize the \texttt{SplittingScheme}. This component is crucial because it subdefines the internal time grids as it is explained in Section~\ref{sec:substep_calculation}. After the temporal hierarchy of operators is established, the scheme instance is returned to the orchestrator.

Finally, the \texttt{Solver} uses the configuration data to initialize the required \texttt{SubSolver} instances; such as advection, diffusion or loss solvers. Once all subsolvers are instantiated and their internal states are prepared, the simulation is ready for execution.

After this initialization pipeline has been completed, the time integration phase encompasses the actual numerical solution of the transport. It begins when a script explicitly demands the calculation's execution by any of the available methods. 

To monitor the progress of the simulation, the \texttt{Solver} first commands the \texttt{cli} module to create and initialize a progress bar. The execution then enters the main temporal loop, iterating over the global time grid. At the beginning of each global timestep, the \texttt{Solver} updates the progress bar and delegates the mathematical advancement of the system by calling the \texttt{SplittingScheme} to apply the splitting step.

Hence, \texttt{SplittingScheme} takes control of the local time advancement. It loops over the initialized \texttt{SubSolver} instances in the mathematically prescribed order. For each operator, the scheme requests the \texttt{SubSolver} to advance the \texttt{State}. The \texttt{SubSolver} then executes its specific numerical routine, all explained in Section~\ref{sec:numerical_operators}. During this process, the \texttt{SubSolver} operates directly on the \texttt{State} object, updating the distribution function in place. 

When all operators have been applied sequentially, the splitting step for the current global time interval is marked as completed. This cyclic process continues until the final simulation time is reached. Upon completion, the \texttt{Solver} stops the \texttt{cli} progress bar and returns the final, fully updated \texttt{State}. 

This hierarchical execution model is justified by the need for numerical stability and code extensibility. By isolating the time-stepping logic from the specific numerical implementations, the architecture allows researchers to seamlessly interchange numerical methods or add new physical operators without modifying the core temporal loop.
\section{Loss mechanisms implemented in \texttt{SAETASS}}
\label{app:loss_functions_implemented}

Among other modules, the \texttt{utils} sub-package of \texttt{SAETASS} contains \texttt{energy\_losses.py}, a comprehensive suite for calculating the energy loss rates and characteristic timescales for various astrophysical processes. These calculations can be used by the user to easily pre-configure solvers and establish the physical regime of the simulation, without having the need of implementing the calculations themselves.

The module evaluates the energy loss rate, $\mathrm{d}E/\mathrm{d}t = b(r,E)$, and the corresponding cooling timescale, $\tau_\mathrm{loss} = E / b(r,E)$, for both hadronic and leptonic populations. The following mechanisms are implemented:

\begin{description}
    \item[Hadronic Losses:] Includes \textit{pion production}, \textit{ionization} and \textit{Coulomb scattering}.
    \item[Leptonic Losses:] Includes \textit{ionization}, \textit{Coulomb scattering}, \textit{Bremsstrahlung}, \textit{synchrotron radiation} and \textit{Inverse Compton scattering}.
\end{description}

The specific mathematical expressions considered for each of the loss mechanisms, concrete references \citep{BlumenthalGould1970, MannheimSchlickeiser1994, KrakauSchlickeiser2015, Ginzburg1979, Evoli2017, Strong1998} and more detailed information on specificities of the module can be found in \texttt{SAETASS} documentation.

\begin{figure}[ht!]
    \centering
    \begin{subfigure}[t]{\textwidth}
        \centering
        \includegraphics[width=0.85\textwidth]{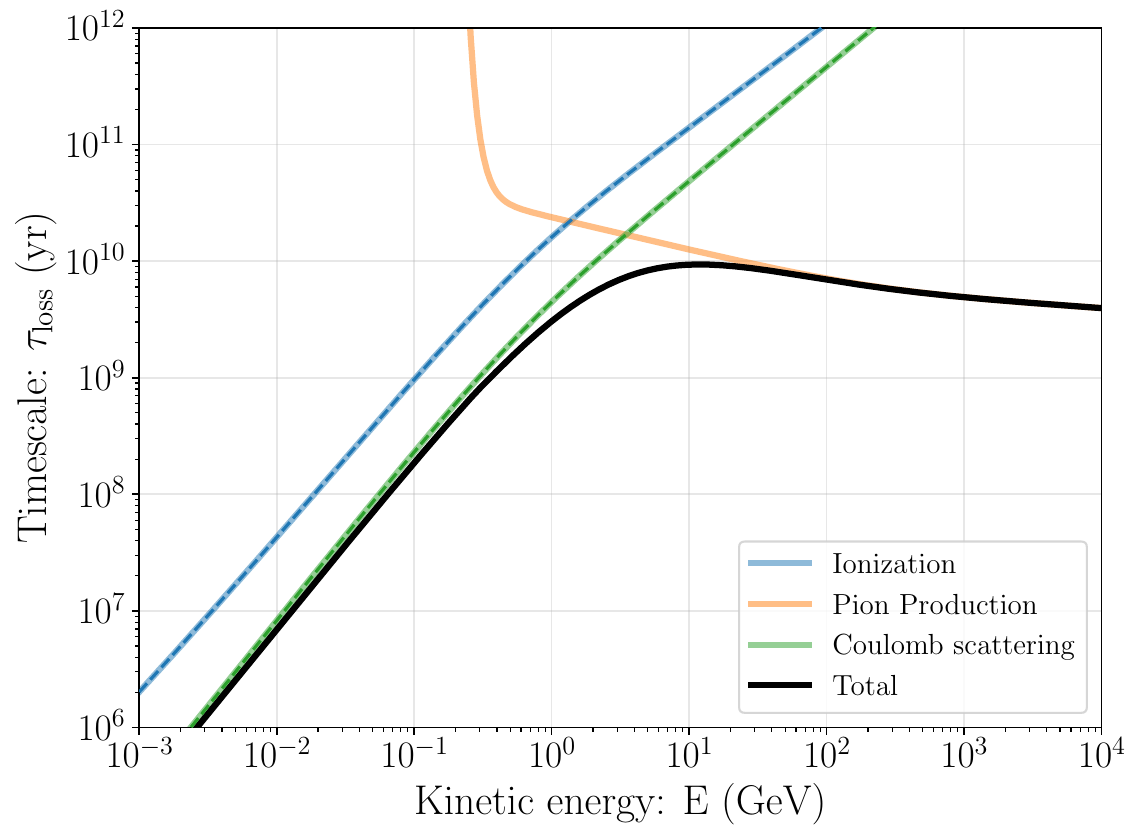}
        \caption{Proton loss timescales.}
        \label{fig:proton_losses}
    \end{subfigure}

    \begin{subfigure}[t]{\textwidth}
        \centering
        \includegraphics[width=0.85\textwidth]{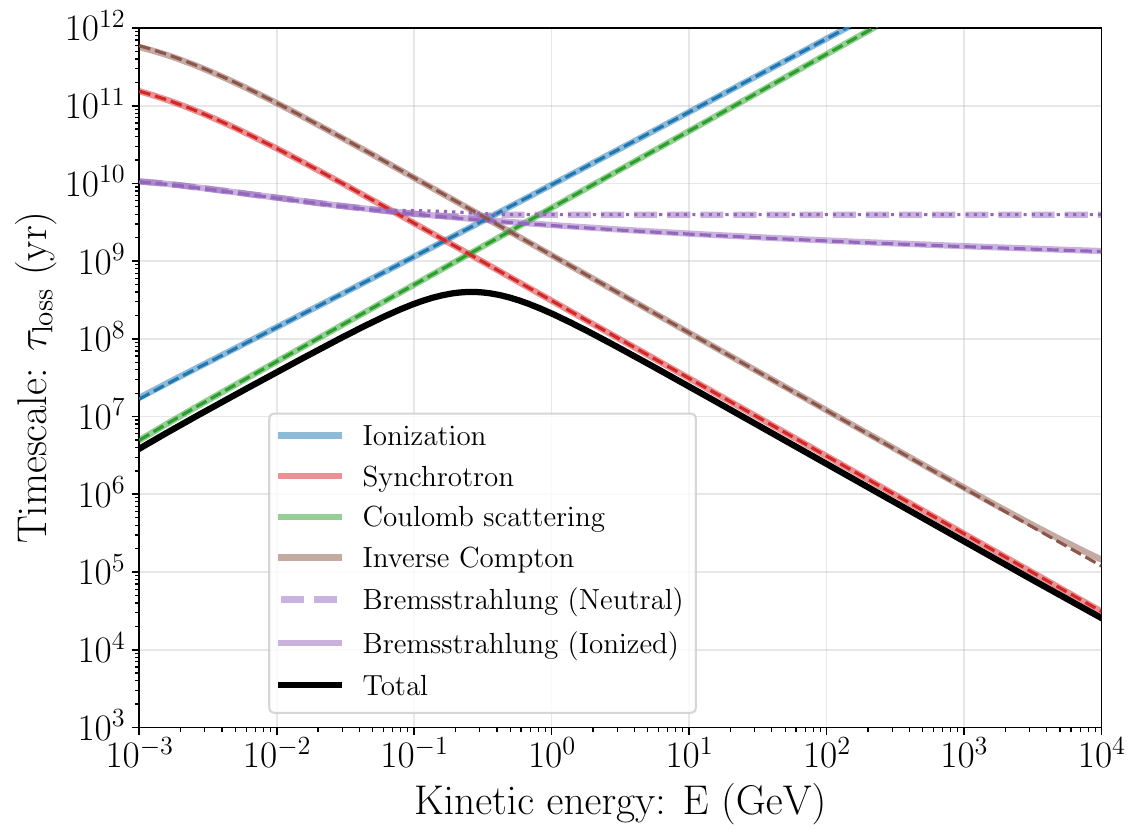}
        \caption{Electron loss timescales.}
        \label{fig:electron_losses}
    \end{subfigure}
    \caption{Cooling timescales, $\tau_\mathrm{loss}$, as a function of particle kinetic energy, $E$. Thick, solid lines represent individual mechanisms calculated via \texttt{SAETASS}; thin, dashed lines represent the reference values from \citep{Strong1998}. The solid black line indicates the total cooling timescale. For detailed information on the specific physical parameters chosen, refer to the corresponding tutorial in \texttt{SAETASS} repository or in its documentation.}
    \label{fig:loss_comparison}
    \vspace{-2cm}
\end{figure}

To validate the implementation, Figure~\ref{fig:loss_comparison} illustrates the cooling timescales for protons and electrons in a standard astrophysical environment. These calculations can be reproduced using the energy losses tutorial inside \texttt{/tutorials} suite in \texttt{SAETASS} directory. The results are compared against the analytical expressions provided by \citep{Strong1998}.

In Figure~\ref{fig:loss_comparison}, \texttt{SAETASS} results are represented by thick, solid lines, with the total loss rate (the sum of all active mechanisms) indicated by a black curve. The reference values from \citep{Strong1998} are plotted as dashed, thinner lines, using matching colours to allow for direct visual comparison without obscuring the primary results. Note that for inverse Compton scattering, the comparison reference corresponds to the simplified Thomson limit expression \citep{Ginzburg1979}, not the expression given by \citep{Strong1998}. However, \texttt{SAETASS} calculation employs the full Klein-Nishina treatment.
\section{Analytical solutions for theoretical validation}
\label{app:analytical}

This appendix collects the analytical results and theoretical arguments used to validate the numerical implementation of the transport equation in spherical symmetry in Section~\ref{sec:tests}. The purpose of this material is: first, to provide closed-form or semi-analytical solutions against which the numerical results can be directly compared; and second, to clarify the expected qualitative and quantitative behaviour of the solutions in the different regimes explored in the validation tests.

\subsection{Pure radial advection with constant velocity}
\label{app:advection_constant}

We consider the transport equation in spherical symmetry restricted to a pure advection process with constant radial velocity. In the absence of diffusion, sources and loss terms, the equation for a scalar quantity $f(t,r)$ reads
\begin{equation}
\frac{\partial f}{\partial t}
+ \frac{1}{r^{2}} \frac{\partial}{\partial r} \left( r^{2} u\, f \right) = 0,
\label{eq:advection_spherical}
\end{equation}
where $u$ is a constant radial velocity. This equation expresses the conservation of $f$ under radial transport in a spherically symmetric geometry.

Expanding the divergence term, the equation~\eqref{eq:advection_spherical} can be rewritten as
\begin{equation*}
\frac{\partial f}{\partial t}
+ u \frac{\partial f}{\partial r}
+ \frac{2u}{r} f = 0 .
\label{eq:advection_expanded}
\end{equation*}
This form makes explicit the geometrical dilution term, which is responsible for the decrease in amplitude of an advected profile as it moves radially outward.

The equation can be solved using the very standard method of characteristics \cite{Courant1989}, which ultimately yields
\begin{equation}
f(r,t) = f_0(r - ut)\,
\left( \frac{r - ut}{r} \right)^{2} ,
\label{eq:advection_solution}
\end{equation}
where $f_0(r)$ denotes the initial condition.

According to the equation~\eqref{eq:advection_solution}, the profile is advected rigidly at speed $u$ while its amplitude decreases as $r^{-2}$ due to spherical geometrical effects. This analytical solution provides a setup for testing the numerical treatment of the advection operator in spherical symmetry, including both the correct propagation speed and the geometrically induced amplitude scaling. This result is used in Section~\ref{sec:adv_validation}.

Although this derivation does not include a source term, the most basic physical intuition derived from the conservation law leads to a straightforward deduction of the steady-state behaviour under continuous injection. Specifically, if we consider a point-like source term at the origin, $Q(r) = Q_0 \delta(r)$, the requirement for a constant mass or particle flux across any spherical shell in a steady-state regime ($\frac{\partial f_\mathrm{ss}}{\partial t}\to 0$) implies that the quantity $r^2u f$ must be spatially invariant. Given a constant advection velocity, the distribution function $f(r)$ must necessarily scale as $r^{-2}$ to compensate for the increasing surface area of the expanding spherical fronts. This result will also be studied in Section~\ref{sec:adv_validation}.

\subsection{Steady-state solution for advection equation with variable coefficient}
\label{app:advection_variable_regularized}

To validate the advection operator in a typical spatially-varying advection astrophysical scenario, we consider a $u(r)\propto r^{-2}$ velocity profile and a localized source $Q(r)$ in the steady-state regime. The transport equation under these conditions reads
\begin{equation}
\frac{1}{r^{2}} \frac{\partial}{\partial r} \left( r^{2} u(r) f_\mathrm{ss} \right) = Q(r).
\label{eq:adv_reg_base}
\end{equation}

We consider a constant velocity profile until $r_\mathrm{c}$ and decreasing from that radius on, and a source term $Q(r)$ localized in $[r_a, r_b]$, where $r_c\ll r_a<r_b$. More precisely, we consider 
\begin{equation*}
u(r) = \begin{cases} 
      u_\mathrm{c} & \text{if } r < r_\mathrm{c}, \\
      u_\mathrm{c} \left( \frac{r_\mathrm{c}}{r} \right)^2 & \text{if } r \ge r_\mathrm{c}
   \end{cases}
   \qquad\text{and}\qquad
Q(r) = \begin{cases} 
      Q_\mathrm{0} & \text{if }r\in[r_a,r_b], \\
      0 & \text{otherwise.}
   \end{cases}
\end{equation*}
In these conditions, equation~\eqref{eq:adv_reg_base} can be integrated to find
\begin{equation*}
r^2u(r)f_\mathrm{ss}(r) = \int_{0}^{r} x^2 Q(x) \, \mathrm{d}x = \begin{cases} 
      0 & \text{if }r < r_a, \\
      \frac{Q_0}{3} \left( r^3 - r_a^3 \right) & \text{if }r_a \le r \le r_b, \\
      \frac{Q_0}{3} \left( r_b^3 - r_a^3 \right) & \text{if }r > r_b.
   \end{cases}
\end{equation*}

The analytical solution for the distribution function is then 
\begin{equation*}
f_\mathrm{ss}(r) =\begin{cases} 
    0 & \text{if }r < r_a, \\
    \frac{Q_0}{3u_\mathrm{c}r_\mathrm{c}^2} \left( r^3 - r_a^3 \right) & \text{if }r_a \le r \le r_b, \\
    \frac{Q_0}{3u_\mathrm{c}r_\mathrm{c}^2}(r_b^3 - r_a^3) & \text{if }r > r_b.
   \end{cases}
\end{equation*}
Specifically, for the region $r >r_b$, the spatial dependence cancels out exactly, leading to a constant plateau. This piecewise solution allows validation of the advection operator with variable velocity, which is studied in Section~\ref{sec:adv_validation}.

\subsection{Pure radial diffusion with constant coefficient}
\label{app:diffusion_constant}

We consider the diffusion equation in spherical symmetry with a constant diffusion coefficient $D$ and in the absence of sources or losses,
\begin{equation*}
\frac{\partial f}{\partial t} = D \left[ \frac{\partial^2 f}{\partial r^2} + \frac{2}{r} \frac{\partial f}{\partial r} \right] ,
\label{eq:diffusion_spherical}
\end{equation*}
where $f(t,r)$ denotes the scalar field of interest. 

For the validation test, we choose the initial condition
\begin{equation}
\label{eq:initial_condition_diff_valid}
f_0(r) = \frac{\pi}{2} \, \mathrm{sinc}(r) = \frac{\pi}{2} \frac{\sin r}{r},
\end{equation}
and we impose regularity at the origin, $\left. \frac{\partial f}{\partial r} \right|_{r=0} = 0$, and $f(t,r_\mathrm{end})=0$ at the outmost boundary.

The function in \eqref{eq:initial_condition_diff_valid} is an eigenfunction of the spherical Laplacian under the present boundary conditions, which allows for a simple analytical solution. The evolution of this profile under constant diffusion is, hence, purely exponential:
\begin{equation*}
f(r,t) = f_0(r) \, e^{ - \pi^2 D t }.
\label{eq:diffusion_solution}
\end{equation*}
Comparison of the numerical solution with this analytical expression allows verification of both the temporal decay and the spatial shape of the profile, as well as convergence behaviour with increasing resolution. This is performed in Section~\ref{sect:diff_validation}.

\subsection{Steady-state solution for diffusion equation with variable coefficient}
\label{app:diffusion_variable_steady}

We consider the diffusion equation in spherical symmetry with spatially dependant coefficient, $D(r)$, and a source term, $Q(r)$. In particular, we are interested in the steady-state limit, which yields
\begin{equation}
\label{eq:diff_variable_coeff}
\frac{1}{r^{2}}\frac{\partial}{\partial r}\left(r^{2} D(r) \frac{\partial f_\mathrm{ss}}{\partial r}\right) = -Q(r).
\end{equation}
To integrate the equation, we define the diffusion coefficient and the source term as,
\begin{equation*}
    \label{eq:definition_coefficients_diff_variable}
    D(r) = D_0(\epsilon+r)^2 \qquad \text{and} \qquad Q(r) = Q_0 r,
\end{equation*}
where $\epsilon>0$ is a \emph{small} parameter introduced to prevent singular behaviour at the origin.

Substituting \eqref{eq:definition_coefficients_diff_variable} into \eqref{eq:diff_variable_coeff} and performing a first integration with respect to $r$ leads to
\begin{equation*}
D_0 r^2 (\epsilon + r)^2 \frac{\partial f_\mathrm{ss}}{\partial r} = -Q_0 \frac{r^4}{4} + C_1.
\end{equation*}
We now impose the regularity boundary condition at the origin, $\left. \frac{\partial f_\mathrm{ss}}{\partial r} \right|_{r=0} = 0$, and the integration constant, $C_1$, vanishes. We have, hence, obtained the gradient of the solution,
\begin{equation*}
\frac{\partial f_\mathrm{ss}}{\partial r} = -\frac{Q_0}{4 D_0} \frac{r^2}{(\epsilon + r)^2}.
\end{equation*}
Integrating once more with respect to $r$ leads to the general form of the solution. Finally, applying the boundary condition at the outer edge, $f_\mathrm{ss}(r_\mathrm{end}) = 0$, to determine the new integration constant, one arrives to a particular analytical solution,
\begin{equation}
\label{eq:steady_state_variable_sol_diff}
f_\mathrm{ss}(r) = \frac{Q_0}{4 D_0} \left[ (r_\mathrm{end} - r) - 2\epsilon \ln\left(\frac{\epsilon + r_\mathrm{end}}{\epsilon + r}\right) - \epsilon^2 \left( \frac{1}{\epsilon + r_\mathrm{end}} - \frac{1}{\epsilon + r} \right) \right].
\end{equation}

The expression \eqref{eq:steady_state_variable_sol_diff} allows for validation of the diffusion operator with variable coefficient. This is done to ensure the numerical convergence towards the steady state solution in Section~\ref{sect:diff_validation}.

\subsection{Steady-state solution for losses equation with a source term}
\label{app:loss_steady}

We consider the transport equation for the momentum distribution $f(t,p)$ including a source term $Q(p)$ and momentum losses given by the function $b(p)$,
\begin{equation}
\frac{\partial f}{\partial t} = Q(p) - \frac{\partial}{\partial p} \left[ b(p) \, f(t,p) \right].
\label{eq:loss_equation}
\end{equation}

For the validation, the source and loss terms are chosen such that the system admits a simple analytical steady-state solution that we obtained from \citep{Evoli2017}. Specifically, the source term is parametrized as a power-law in momentum with index $\alpha$, and the loss term as a power-law with index $\beta$, while the momentum range of characteristic value $p_0$ extends up to $p_\mathrm{end}$. This is, we are considering
\begin{gather*}
    Q(p) = Q_0 \left( \frac{p}{p_0} \right)^{-\alpha}\qquad\text{and}\qquad b(p) = -b_0 \left( \frac{p}{p_0} \right)^{\beta},
\end{gather*}
with $Q_0>0$, $b_0>0$. We restrict our validation test to the physically relevant regime where $\alpha>\beta>1$ (strictly requiring $\alpha\neq 1$ to avoid a logarithmic solution), which represents typical astrophysical injection and radiative loss spectra.

In the steady-state limit, $\frac{\partial f_\mathrm{ss}}{\partial t}\to 0$, and the equation~\eqref{eq:loss_equation} can be integrated analytically to yield
\begin{equation}
f_\mathrm{ss}(p) = \frac{Q_0 \, p_0}{(1-\alpha) \, b_0} \left[ \left( \frac{p_\mathrm{end}}{p_0} \right)^{1-\alpha} - \left( \frac{p}{p_0} \right)^{1-\alpha} \right] \left( \frac{p}{p_0} \right)^{-\beta} ,
\label{eq:loss_steady_state}
\end{equation}
where $p_\mathrm{end}$ represents the upper boundary of the momentum domain. 

This setup allows for validation for the loss operator, which is performed observing that the numerical solution converges in time towards the solution \eqref{eq:loss_steady_state}. This analysis is done in Section~\ref{sec:loss_validation}.
\section{Validation via method of manufactured solutions}
\label{app:mms_validation}

To rigorously test the numerical implementation of the time-dependant transport operators in spherical symmetry, we employ the method of manufactured solutions (MMS). This approach allows us to validate the code against complex scenarios with simultaneous spatial and temporal dependencies by introducing a source term $Q_\text{MMS}(t,r)$ that ensures a prescribed function $f_{\text{MMS}}(t,r)$ is a solution to the PDE. To ensure physical consistency, we choose target solutions that are strictly non-negative.

\subsection{Advection with space and time dependence}
\label{app:advection_mms}

We consider the advection equation in the presence of a source term,
\begin{equation}
\label{eq:mms_eq_adv}
\frac{\partial f}{\partial t} + \frac{1}{r^{2}} \frac{\partial}{\partial r} \left( r^{2} u(t,r) f \right) = Q(t,r).
\end{equation}
We define the manufactured solution and the velocity field as
\begin{gather*}
    f_{\text{MMS}}(r,t) = (2 + \cos(t)) e^{-r}\qquad\text{and}\qquad u(r,t) = \frac{r}{1+t}.
\end{gather*}

Substituting in \eqref{eq:mms_eq_adv} leads to a required source term $Q_{\text{MMS}}$ which takes the form
\begin{equation}
Q_{\text{MMS}}(r, t) = e^{-r} \left[  \frac{(2+\cos(t))(3-r)}{1+t}-\sin(t) \right].
\end{equation}

This test specifically probes the solver's ability to handle the geometric divergence term coupled with a time-varying, non-uniform velocity field while maintaining the positivity of the distribution. Such results are shown in Section~\ref{sec:time_validation}.

\subsection{Diffusion with non-stationary coefficient.}
\label{app:diffusion_mms}

For the diffusion operator validation, we consider the case of a time varying diffusion coefficient, hence having the equation
\begin{equation}
\label{eq:mms_diff_eq}
\frac{\partial f}{\partial t} = \frac{1}{r^{2}} \frac{\partial}{\partial r} \left( r^{2} D(t) \frac{\partial f}{\partial r} \right) + Q(r, t).
\end{equation}
We use as target solution and diffusion coefficient 
\begin{equation*}
   f _{\text{MMS}}(r,t) = (2 + \cos(t)) e^{-r^2}\qquad\text{and}\qquad D(t) = D_0(1+t).
\end{equation*}

Again, substituing in the equation~\eqref{eq:mms_diff_eq}, leads to the need source term, this is,
\begin{equation}
Q_{\text{MMS}}(r, t) = e^{-r^2} \left[  2 D_0 (1+t) (2 + \cos(t)) (3 - 2r^2)-\sin(t) \right].
\end{equation}

The comparison between the numerical evolution and $f_{\text{MMS}}$ provides a direct measure of the discretization error for the radial diffusion operator under non-steady conditions. Such analysis is performed in Section~\ref{sec:time_validation}.

\acknowledgments

We are grateful to Carmelo Evoli and Iurii Sushch for valuable discussions and comments during the paper drafting.

Authors acknowledge financial support from the Severo Ochoa grant CEX2021-001131-S funded by MCIN/AEI/ 10.13039/501100011033. Authors also acknowledge financial support from the Spanish ``Ministerio de Ciencia e Innovaci\'on'' through grant PID2022-139117NB-C44 and grant CNS2023-144504 funded by MICIU/AEI/ 10.13039/501100011033 and by the European Union NextGenerationEU/PRTR.  J.M. G.M. acknowledges financial support from the FPU grant FPU24/01674 funded by the Spanish ``Ministerio de Ciencia, Innovaci\'on y Universidades''.

\bibliographystyle{JHEP} 
\bibliography{references}

\end{document}